\documentclass[journal,12p]{IEEEtran}
\usepackage{amsmath,amssymb}
\usepackage{latexsym}
\usepackage{graphicx,color}

\newsavebox{\trellis}
\newsavebox{\trellislines}
\newsavebox{\trellisdots}
\newsavebox{\couple}

\newtheorem{lemma}{Lemma}
\newtheorem{corollary}{Corollary}

\def\R{\mathbb R}
\def\Z{\mathbb Z}
\def\supremum{\mathop{\rm sup}}
\def\infimum{\mathop{\rm inf}}
\def\dbar{\overline{d}}
\def\R{\mathbb{R}}
\def\qed{\hfill $\Box$}

\definecolor{mathcolor}{rgb}{0.2,0.6,0.2}

\begin{document}
\title{
%{\large \today\\ \color{red}  DRAFT\\
%Do not circulate without permission}\\
Rate-Constrained Simulation\\ and Source Coding IID Sources}

\author{Mark Z. Mao,~\IEEEmembership{Member,~IEEE,}
        Robert M. Gray,~\IEEEmembership{Life Fellow,~IEEE},
        and Tamas Linder,~\IEEEmembership{Senior Member,~IEEE}%
\thanks{M. Z. Mao and R. M. Gray are with the Department
of Electrical Engineering, Stanford University, Stanford,
CA, 94305, USA. Tamas Linder is with the Department of Mathematics and Statistics,
Queens University, Kingston, Ontario, Canada~K7L 3N6,
e-mail: markmao at stanford.edu, rmgray at stanford.edu,
linder at mast.queensu.ca.}% <-this % stops a space
\thanks{Research partially supported by the NSF under grant CCF-0846199-000
and by the Natural Sciences and Engineering Research
Council (NSERC) of Canada.}% <-this % stops a space
\thanks{\today.}
}

%\markboth{IEEE Transactions on Information Theory}%
%{Mao and Gray}


\maketitle

\begin{abstract}

Necessary conditions for asymptotically optimal sliding-block or stationary codes
for source coding and  rate-constrained simulation of memoryless sources are presented and used
to motivate a design technique for  trellis-encoded source coding
and rate-constrained simulation. The code structure has intuitive
similarities to classic random coding arguments as well as to  ``fake process'' methods and 
alphabet-constrained methods. 
Experimental evidence shows that the approach provides comparable
or superior performance in comparison with 
previously published methods on common examples, 
sometimes by significant margins. 
%Unlike previous methods,
%the performance has not yet been found to hit a plateau 
%as code complexity grows.

\end{abstract}


\section{Introduction}
\IEEEPARstart{T}{he} basic goal of Shannon source coding with a
fidelity criterion or lossy data compression is to covert an
information source $\{X_n\}$ into bits which can be decoded into a
good reproduction of the original source, ideally the best possible
reproduction with respect to a fidelity criterion given a constraint
on the rate of transmitted bits.  Memoryless discrete-time sources
have long been a standard benchmark for testing source coding or data
compression systems. Although of limited interest as a model for real
world signals, independent identically distributed (IID) sources
provide useful comparisons among different coding methods and
designs. In addition, specific examples such as Gaussian and uniform
sources can provide intuitive interpretations of how coding schemes
yield good performance and they can serve as building blocks for more
complicated processes such as linear models driven by IID processes.

A separate, but intimately related, topic is that of rate-constrained
simulation --- given a ``target'' random process such as an IID
Gaussian process, what is the \emph{best} possible imitation of the
process that can be generated by coding a simple discrete IID process
with a given (finite) entropy rate?  Here ``best'' can be quantified by a
metric on random processes such as the generalized Ornstein
$\overline{d}$ distance (or Monge-Kantorovich transportation distance/
Wasserstein distance extended to random processes).  For example, what
is the best imitation Gaussian process with only one bit per symbol?

%% Here and throughout the paper the emphasis is on the special case of
%% integer rates $R$ and, in particular, on the case of $R=1$ bit per
%% symbol since this provides the simplest possible context and notation
%% for the basic ideas and examples.  In particular, in the source coding
%% system one bit is communicated for each sample $X_n$ and in the
%% simulation system one simulated sample $\tilde{X}_n$ is generated for
%% each coin flip.

Intuitively and mathematically \cite{Gray:08,GrayL:09}, if the source code
is working well, one would expect the channel bits produced by the source
encoder to be approximately IID and the resulting reproduction process to be
as close to the source as possible with a one bit per symbol channel. Thus
the decoder driven by coin flips should produce a nearly optimal simulation.
Conversely, if an IID source driving a stationary code produces a good
simulation of a source, the code should provide a good decoder %encoder 
in a source coding system with an encoder matching possible decoder outputs to the source sequence,
e.g., a Viterbi algorithm.

Rigorous results along this line were developed in \cite{Gray:77},
showing that the two optimization problems are equivalent and optimal
(or nearly optimal) source coders imply optimal (or nearly optimal)
simulators and vice versa for the specific case of 
stationary codes  %(shifting the input results in a corresponding shift of the output)  
and  sources that are $B$-processes (stationary codings of 
IID processes).

Results that are similar in spirit were developed for more general
sources by Steinberg and Verdu \cite{SteinbergV:96}, where
other deep connections between process simulation and rate-distortion theory
were also explored.  However, results in \cite{SteinbergV:96} are for
asymptotically long block codes while our focus is on stationary codes
--- especially on stationary decoders of modest memory ---
and on the behavior of processes rather than on the asymptotics of
finite-dimensional distributions, which might not correspond to the
joint distributions of a stationary process.

%% If the goal is to approximate one process by another, then the
%% approximating process should 
%% inherit key statistical properties of the
%% original, including stationarity, ergodicity, and mixing
%% properties. This cannot be accomplished using block codes, even if
%% they are ``stationarized'' by introducing a uniformly distributed
%% random starting point for the blocking.  The random time origin indeed
%% gives a stationary process, but in general ergodicity and mixing
%% properties are lost and periodicities and other blocking artifacts are
%% introduced.

%In this paper, we 
We introduce a design technique for trellis-encoded
source coding based on designing a stationary decoder to
approximately satisfy necessary conditions for optimality 
(analogous to the Lloyd
algorithm for vector quantizer design~\cite{GershoG:92}) and using a matched
Viterbi algorithm as an encoder (analogous to the minimum distortion
encoder in the Lloyd algorithm). The combination of a good decoder with a
matched search algorithm as the encoder is the most common implementation
of trellis source codes. 
Previous
work
\cite{%Wilson:75,
WilsonL:77,Pearlman:82,vanderVleutenW:95,ErikssonAG:07} for 
trellis encoding system design has been based largely on
intuitive guidelines, assumptions, or formal axioms for good code design.
In contrast, we prove 
several necessary conditions which optimal or asymptotically
optimal source codes must satisfy, including some properties simply
assumed in the past. Examples of such properties are
Pearlman's
observations \cite{Pearlman:82} that the marginal reproduction
distribution should approximate the Shannon optimal reproduction and
that the reproduction process should be approximately white. We give a code
construction %%??involving no random selection 
which provably satisfies a
key necessary condition and which is shown experimentally to satisfy
the other necessary conditions while providing performance comparable
to or superior to previously published work, and in many cases,
remarkably close to the theoretical limit.

\iffalse %said elsewhere
In particular, the
performance of our trellis source encoder has not yet reached a
plateau and seems to 
approach the Shannon limit with increasing encoder complexity. 
\fi

The rest of the paper is organized as follows. In
Section~\ref{sect:prelim} we give an overview of definitions and
concepts we need for stating our results and in
Section~\ref{sect:necessary} we state and prove the necessary conditions
  for optimum trellis-encoded source code
  design. Section~\ref{sect:algorithm} introduces the new design
  technique and Section~\ref{sect:numerical} presents experimental
  results for encoding memoryless Gaussian, uniform, and Laplacian 
  sources.

\section{Preliminaries} 
\label{sect:prelim}

\subsection{A note on notation} 
%\emph{Notation}

We deal with random objects which will be denoted by
capital letters. These include random variables $X_n$, $N$-dimensional random vectors
$X^N=(X_0,X_1,\ldots , X_{N-1})$, and random processes $\{X_n; n\in \Z \}$,
where $\Z$ is the set of all integers. The generic notation
$X$ might stand for any of these random objects, where the specific nature will either be
clear from context or stated (this is to avoid notational clutter when possible). Lower case letters
will correspond to sample values of random objects. For example, given an alphabet $A$
(such as the real line $\R$ or the binary alphabet $\{0,1\}$), then a random variable $X_n$ 
may take on values $x_n\in A$, an $N$-dimensional random vector $X^N$ may take on values 
$x^N\in A^N$, the Cartesian
product space, and a random process $\{X_n; n\in \Z \}$ may take on values 
$\{x_n; n\in \Z \}=(\cdots , x_{-1} , x_0 , x_{1}, \cdots )\in
A^\infty$. A lower case
letter without subscript or superscript may stand for a member of any of these spaces, depending
on context. 

\subsection{Stationary and sliding-block codes}
%\emph{Stationary and sliding-block codes}

A stationary or sliding-block code is  a time-invariant filter, in general nonlinear. It
operates on an
input sequence to produce an output sequence in such a way that shifting the input sequence
results in a shifted output sequence. More precisely, a stationary code 
$\bar{f}$ with an input alphabet
$A$ (typically  $\R$ or a Borel subset for an encoder
or  $\{0,1\}$ for a decoder) and output alphabet $B$ (typically $\{0,1\}$ for an encoder  
or some subset $\R$ for a decoder) is a measurable mapping (with respect to 
suitable $\sigma$-fields)
of an infinite input sequence (in $A^\infty$) into an infinite output sequence (in $B^\infty$) with
the property that $\bar{f}(T_Ax)=T_B\bar{f}(x)$, where $T_A$ is the (left) shift on $A^\infty$, that is,
$T_A(\cdots , x_{-1} , x_0 , x_{1}, \cdots )=( \cdots , x_{0} , x_{1}, x_{2} \cdots )$. The 
sequence-to-sequence mapping
from $\bar{f}:A^\infty \rightarrow B^\infty$
is described by the sequence-to-symbol mapping defined by code
output at time 0, $f(x)=\bar{f}(x)_0$ since $\bar{f}(x)_n=\bar{f}(T_A^nx)_0=f(T_A^nx)$. 
More concretely,
the sequence-to-symbol mapping $f$ usually depends on only a finite window of the data, in which
case the output random process, say $\{Y_n\}$, can be expressed as
$
Y_n=f(X_{n-N_1}, \cdots , X_n, \cdots , X_{n+N_2})
$,
a mapping on the contents of a shift register containing
$L=N_1+N_2+1$ samples of
the input random process $\{X_n\}$.
Both %the sequence-to-symbol mapping 
$f$ and %the equivalent sequence-to-sequence mapping 
$\bar{f}$ will be referred to as
stationary or sliding-block codes.
%%the former name reflecting the fact that a stationary input and a
%% stationary code yield a jointly stationary input/output process and
%% the latter name reflecting the contrast with traditional block code
%% operation --- each output depends on overlapping blocks which are
%% ``slid'' into the shift register one symbol at a time (when $R=1$).
\iffalse
Stationary codes (unlike traditional block codes) are well-defined for
infinite length codes, that is, codes are well-defined for which 
an infinite  input sequence is viewed to produce a single output
symbol. %This is useful for theory and interpretation.
\fi

Unlike block codes,
stationary codes preserve statistical characteristics of the
coded process, including stationarity, ergodicity, and mixing. 
If a stationary 
and ergodic source $\{X_n\}$
is encoded into bits by a stationary code $f$, which are  in turn decoded into a reproduction
process $\{\hat{X}_n\}$ by another stationary code $g$, then the resulting pair process
$\{X_n,\hat{X}_n\}$ and output process $\{\hat{X}_n\}$ are also stationary and ergodic.

Given any block code, a stationary code with similar properties can be constructed (at least
in theory) and vice versa. Thus good codes of one type can be used to construct good codes
of the other (at least in theory) and the optimal performance for the two classes of codes is the same~\cite{Shields:79,GrayNO:75,Gray:75,Gray:90}. 
\iffalse
Far more is known about implementing
block codes than stationary codes. This paper is intended as a contribution to the 
the design of stationary codes and to trellis encoding source coders, which match a search algorithm
to a stationary decoder.
\fi

%% \subsection*{$B$-processes} $B$-processes play a central role in
%% ergodic theory because they are the class of processes (discrete or
%% continuous alphabet) for which two processes are isomorphic (one
%% can be mapped into the other in a stationary and invertible, with
%% probability 1, way) if and only if their entropy rates are
%% equal~\cite{Ornstein:73,Ornstein:75}).  The class of $B$-processes
%% includes, for example, all regular (purely nondeterministic)
%% stationary discrete-time Gaussian random
%% processes~\cite{Smorodinsky:71}, all time-invariant linear
%% filterings of IID processes such as moving average and
%% autoregressive processes driven by IID processes~\cite{Gray:09},
%% and all discrete alphabet stationary and ergodic finite order
%% Markov sources and limits of such processes in the Ornstein $\dbar$
%% distance~\cite{Ornstein:73,Ornstein:75}.

\subsection{Fidelity and  distortion}
%\emph{Fidelity and  distortion}

A distortion measure $d(x,y)$, $x\in A, y\in
\hat{A}$ is a nonnegative measurable function (with respect to 
suitable $\sigma$-fields). A fidelity criterion is a family of distortion measures
$d_N(x^N,y^N), x^N \in A^N, y^N\in \hat{A}^N$, $N=1,2,\ldots $. We assume that the  fidelity criterion is additive (or single-letter):
\[
d_N(x^N,y^N)= \sum_{i=0}^{N-1} d(x_i,y_i),
\]
where $d=d_1$. 
Throughout the paper, we make the standard assumption that 
$\hat{A}\subset A$ and $d(x,x)=0$.
Given random vectors $X^N,Y^N$ with a joint distribution $\pi^N$, the
average distortion is defined by the expectation
$
d (\pi^N)= % d_N (\pi^N)= %%drop the _N for consistency with I(\pi^N)
E[d_N(X^N,Y^N)].
$

Given a stationary pair process $\{X_n,Y_n\}$, the average
distortion between $N$-tuples is given by the single-letter
characterization $N^{-1}E[d_N(X^N,Y^N)]%&=&N^{-1}d_N(\pi^N)\\&=&
=E[d(X_0,Y_0)]= d (\pi^1 )$ and hence a measure of the fidelity (or,
rather, lack of fidelity) of a stationary coding and decoding of a
stationary source $X_n$ into a reproduction $\hat{X}_n$ is the average
distortion
$
D(f,g)=E[d(X_0,\hat{X}_0)].
$
The emphasis in this paper will be the case where $A=\R$ and the distortion is the
common squared error distortion, $d(x,y)=(x-y)^2$.
%(so that the distortion on $n$-tuples is the
%square of the $\ell_2$ distance). 
Also of interest is the Hamming distortion, where
$d(x,y)=0$ if $x=y$ and 1 otherwise. 

Throughout the paper  we assume that the stationary process $\{X_n\}$ and the
distortion measure $d$ satisfy the following standard reference letter
condition:  there exists $\hat{x} \in \hat{A}$ such that 
$E [d(X_0,\hat{x})]<\infty$. In particular, when the distortion is the
squared error, we always assume that the source has finite variance. 

\subsection{Optimal source coding}
%\emph{Optimal source coding}

Let $\mathcal{C}(A,B)$ denote the collection of all sliding-block
codes with input alphabet $A$ and finite  output alphabet $B$ 
of size  $\|B\|$.
The operational
distortion-rate function for source $X$ is defined by
\[
\delta_X(R)=\infimum_{f\in \mathcal{C}(A,B),g\in \mathcal{C}(B,\hat{A}):
\|B\|\le 2^R
} D(f,g).
\]
Note that $\delta_X(R)$ is defined for the discrete set of $R$ values
such that $R=\log k$ for some nonnegative integer $k$.

\subsection{Distance measures for random vectors and processes}
%\emph{Distance measures for random vectors and processes}

A distortion measure $d$ induces a natural notion of a ``distance''
between random vectors and processes (the quotes will be removed when
the relation to a true distance or metric is clarified).  
\iffalse
The basic
idea has many names and a wide literature spanning many fields. 
It is
basically the extension to general distortion measures of Ornstein's
$\overline{d}$ (d-bar) distance (which is based on the the Hamming
distortion) or, equivalently, the extension of the classic
Monge/Kantorovich/Wasserstein (or Vasershtein)
distance~\cite{Monge:81,Kantorovich:48,Vasershtein:69},
also called optimal transportation cost, 
from random vectors to random processes. 
\fi
The \emph{optimal transportation cost} between two probability
distributions, say $\mu_X$ and $\mu_Y$, corresponding to random
variables (or vectors) defined on a common (Borel) probability space
$(A,\mathcal{B}(A))$ with a nonnegative cost function $d$ is defined as
$$
\mathcal{T}(\mu_X,\mu_Y)
=
\infimum_{\pi \in \mathcal{P}(\mu_X,\mu_Y)}
E_\pi d(X,Y),
$$
where $\mathcal{P}(\mu_X,\mu_Y)$ is the class of all probability distributions
on $(A,\mathcal{B}(A))^2$ having $\mu_X$ and $\mu_Y$ as marginals, that is,
$\pi (F\times A)=\mu_X(F)$,
$\pi (A\times F )=\mu_Y(F)
$ for all $F\in \mathcal{B}(A)$. 
\iffalse %no longer need
Any $\pi$ having the prescribed
marginals $\mu_X$ and $\mu_Y$ is called a \emph{coupling} of $\mu_X$
and $\mu_Y$. 
\fi
The reader is referred to
Villani~\cite{%Villani:03,
Villani:09} and Rachev and
R\"{u}schendorf~\cite{RachevR:98} for extensive development and
references. The most important special case  is when the cost function is a nonnegative power of an
underlying metric:
$%\begin{equation}\label{eq:metric}
d(x,y)=m(x,y)^r
$, %\end{equation}
where $A$ is a complete, separable metric (Polish) space with respect
to $m$. In this case $\mathcal{T}(\mu_X,\mu_Y)^{\min (1,1/r)}$ is a
metric. 
%The case $r=0$ is used to denote the Hamming distance.  
The
notation $\mathcal{T}_2$ and $\mathcal{T}_0$ will be used to denote
the two most important cases of the optimal transportation cost with
respect to the squared error and Hamming distance, respectively.

Given two processes with process distributions $\mu_X$ and $\mu_Y$ on
$(A^{\infty},\mathcal{B}(A^{\infty}))$, let
$\mu_{X^N}$ and
$\mu_{Y^N}$ denote the induced $N$-dimensional distributions
for all positive integers $N$. Let $d_N$ be an additive %single-letter
distortion measure induced by $d(x,y)$, $x,y\in A$. 
Define the (generalized) $\dbar$ distance~\cite{GrayNS:75}
between two stationary processes
%(also called the $\overline{\rho}$ distance~\cite{GrayNS:75})
\begin{eqnarray*}
\overline{d}(\mu_X,\mu_Y)&=&
\supremum_N N^{-1}
\mathcal{T}(\mu_{X^N},\mu_{Y^N}).%\\
%&=&
%\supremum_N N^{-1}
%\infimum_{\pi \in \mathcal{P}(\mu_{X^N},\mu_{Y^N})}
%E_\pi d_N(X^N,Y^N).
\end{eqnarray*}
If $d$ is a metric, then so is $\overline{d}$. If $d$ is the Hamming
metric, this is Ornstein's $d$-bar
distance~\cite{Ornstein:73,Ornstein:75}. 
%%It was generalized to the
%nonmetric example of squared error in
%\cite{GrayNS:75}.%,PapantoniG:79}. See also \cite{Gray:09}.  
If $d$ is a
power of an underlying metric, % as in (\ref{eq:metric}), 
then
$\overline{d}(\mu_X,\mu_Y)^{\min (1,1/r)}$ will also be a metric.  We
will refer to $\overline{d}$ as the ``$\overline{d}$-distance''
whether or not it is actually a true metric.  We distinguish the most
important cases by subscripts, in particular $\overline{d}_2$ denotes
$\overline{d}$ with $d$ squared error (and hence
$\sqrt{\overline{d}_2}$ is a metric) and $\overline{d}_0$ denotes
$\overline{d}$ with $d$ equal to the Hamming distance
($\overline{d}_0$ is a metric). 
%Usually in the literature $\overline{d}_0$ is
%adopted when considering discrete alphabet processes and
%$\overline{d}_2$ when considering continuous alphabet processes.  In
%source coding systems, typically both continuous alphabet and discrete
%alphabet processes are involved and hence both distances may crop up.

\iffalse
The basic definition for the $\overline{d}$ distances is in terms of a supremum over
vector optimizations, but 
\fi
For stationary processes there is a simpler characterization of $\overline{d}$:
\begin{equation}
\overline{d}(\mu_X,\mu_Y)=
\infimum_{\pi \in \mathcal{P}(\mu_X,\mu_Y)}
E_\pi [d(X_0,Y_0)]
\end{equation}
where  the infimum is over all stationary processes (or stationary and ergodic processes
if $\mu_X$ and $\mu_Y$ are ergodic). This and many
other properties of the $\overline{d}$ and generalized $\overline{d}$
are detailed in \cite{Ornstein:73,Ornstein:75,GrayNS:75,Gray:09}. 
Properties relevant here include the following:
\begin{enumerate}
%\item
%The infimum over
%all couplings is  a minimum.
%% Not proved correctly in Gray:09 or earlier edition, not considered
%%in other references. Not needed, remove.
\item For stationary processes,
\begin{equation}
\dbar(\mu_X,\mu_Y) = \lim\limits_{N\to \infty} N^{-1} \mathcal{T}(\mu_{X^N},
  \mu_{Y^N}). \label{eq:sup-limit}
\end{equation}
\item If the processes are both IID, then
\begin{equation}
\dbar (\mu_X,\mu_Y)
=
\mathcal{T}(\mu_{X_0},\mu_{Y_0}).
\end{equation}
\iffalse
and hence the $\dbar$-distance reduces to the transportation distance between the one dimensional
marginal distributions. 
\fi

\item If the processes are both stationary and ergodic, the distance
  can be expressed as the infimum over the limiting distortion between
  any two frequency-typical sequences of the two processes. 
  Thus the $\overline{d}$-distance between the two processes is the
  amount by which a frequency-typical sequence of one process must be
  changed in a time average $d$ sense to produce a frequency-typical
  sequence of another process.

\end{enumerate}

The $\dbar$ process distance  can be used to characterize both the optimal source coding and the optimal
rate-constrained simulation problem.
Let   $\{X_n\}$ be a random process described by a process distribution $\mu_X$ and 
let $\{Z_n\}$ be an IID \emph{equiprobable} random process 
with alphabet $B$ of size  $\|B\|=2^R$ and distribution
$\mu_Z$. The optimal simulation
of the process $X=\{X_n\}$ 
with process distribution $\mu_X$
given the process $Z=\{Z_n\}$ with process distribution $\mu_Z$ 
and reproduction alphabet $\hat{A}$ is characterized by
\begin{equation}
\Delta_{X\mid Z}(R)=
\infimum_{f\in\mathcal{C}(B,\hat{A})} \dbar (\mu_X,\mu_{\bar{f}(Z)})
\end{equation}
where $\mu_{\bar{f}(Z)}=\mu_{Z}
\bar{f}^{-1}$ is the process distribution resulting from a stationary coding 
of $Z$ using $f$, i.e., for all events $F$
$
\mu_{\bar{f}(Z)}(F)=
\mu_{Z}(
\bar{f}^{-1}(F))$.
The notation for $\Delta_{X\mid Z}(R)$ is redundant since $R$ determines
the distribution of $Z$ and vice versa. As in the definition of
the operational rate-distortion function, $R$ is of the form $R=\log
k$ for some nonnegative integer $k$. 
\iffalse
Thus $\Delta_{X\mid Z}(R)$ quantifies 
the best approximation to the target process $X$ in the $\dbar$-distance
over all stationary codings of an IID process $Z$ with an alphabet of size
$2^R$ . 
\fi

%% Constrained-rate simulation can be thought of as a variation on the
%% idea of pseudo-random noise generation. A pseudo-random noise
%% generator typically attempts to produce a good fake of a sequence of
%% uniform (0,1) IID random variables given a sequence of deterministic
%% mappings on an initial (usually random) seed. Here a sequence of coin
%% flips is used to generate a fake of a more random process, such as an
%% IID Gaussian random process. Process distance measures permit the
%% quantification of how successful the fake process is.

\subsection{Entropy rate}
%\emph {Entropy rate}

Alternative characterizations of the optimal source coding and 
simulation performance can be stated in terms of the
entropy rate of a random process. As we will be dealing with both discrete and continuous
alphabet processes and with some borderline processes that have continuous alphabets
yet finite entropy, suitably general notions of entropy as found in mathematical information
theory and ergodic theory are needed (see, e.g..,
\cite{Pinsker:64,Ornstein:73,Ornstein:75,Gray:09}). 
For a finite-alphabet random process,
define as usual the
Shannon entropy of a random vector or, equivalently, of its distribution by
$
H(X^N)=H(\mu_{X^N})=
-\sum_{x^N} \mu_{X^N}(x^N)\log \mu_{X^N}(x^N) $
and  the Shannon entropy rate of the process $X$ by
$
H(X)=H(\mu_X )=\infimum_N N^{-1} H(X^N).$
If the process is stationary, then
\begin{equation}\label{entropyrate}
H(X)=\lim_{N\rightarrow \infty} N^{-1}H(X^N).
\end{equation} 
In the general case of a continuous alphabet,
the entropy rate is given by the Kolmogorov-Sinai invariant 
$
H(X)=\supremum_{f} H(\mu_{\overline{f}(X)})$,
where the supremum is over all finite-alphabet stationary codes. 
\iffalse
In
other words, the entropy rate of a continuous alphabet random process
is the supremum of the ordinary Shannon entropy rates of
finite-alphabet coded versions of the process. 
\fi
It is important to note that
(\ref{entropyrate}) need not hold when the alphabet is not finite and
that a random process with a continuous alphabet can have an infinite
finite-order entropy and a finite entropy rate.

\subsection{Constrained entropy rate optimization}
%\emph{Constrained entropy rate optimization}

A stationary and ergodic process is called a $B$-process if it is
obtained by  a stationary coding of an IID process. 
If the source is stationary and ergodic, then~\cite{Gray:77}
\begin{equation}\label{Deltadef}
\Delta_{X\mid Z}(R)=
\infimum_{B\mbox{\scriptsize-processes } \nu: H(\nu)\le R} \dbar (\mu_X,\nu),
\end{equation}
that is, the best simulation by coding coin flips in a stationary
manner has the same performance as the best simulation of $X$ by any $B$-process having entropy
rate $R$ bit per symbol or less.  If $X$ were itself discrete and a
$B$-process with entropy rate less than or equal to $R$, then
Ornstein's isomorphism theorem \cite{Ornstein:73,Ornstein:75} (or the weaker Sinai-Ornstein theorem)
implies that $\Delta_{X\mid Z}(R)=0$. In words, a $B$-process can be
stationarily encoded into any other $B$ process having equal or
smaller entropy rate.

The $\dbar$-distance also yields a characterization of the operational distortion rate function
~\cite{GrayNO:75}:
\begin{equation}\label{eq:deltadbar}
\delta_X(R)=%\infimum_{f,g} D(f,g)=
\infimum_{\nu: H(\nu )\le R}\dbar (\mu_X,\nu),
\end{equation}
where the infimum is over all stationary and ergodic processes. 
Comparing (\ref{Deltadef}) and (\ref{eq:deltadbar}), obviously
$ \Delta_{X|Z}(R)\ge \delta_X(R). $ If the source
$X$ is also a $B$-process, then the two infima are the same and $
\Delta_{X|Z}(R)= \delta_X(R)$.  

A related operational distortion-rate
function resembling the simulation problem replaces the
encoder/decoder with a common encoder output/decoder input
alphabet by a single code into a
reproduction having a constrained entropy rate. Suppose that a source
$X$ is encoded by a sliding-block code $f$ directly into a
reproduction $\hat{X}$ with process distribution
$\mu_{\hat{X}}=\mu_{\bar{f}(X)}$. What coding yields the smallest
distortion under the constraint that the output entropy rate is less
than or equal to $R$? In this case, unsurprisingly 
\begin{equation}
\infimum_{f\in \mathcal{C}(A,\hat{A}):H(\mu_{\bar{f}(X)})\le
  R}E[d(X_0,\hat{X}_0)] =
\delta_{X}(R).
\end{equation}
These relations implicitly define optimal codes and optimal performance,
but they do not say how to evaluate the optimal performance or design the codes
for a particular source. The Shannon rate-distortion function solves the first problem.

\subsection{Shannon rate-distortion functions}
%\emph{Shannon rate-distortion functions}

In the discrete alphabet case the $N$th order average mutual information between random vectors
$X^N$ and $Y^N$ is given by $I(X^N,Y^N)=H(X^N)+H(Y^N)-H(X^N,Y^N)$. In general
$I(X^N,Y^N)$ is given as the supremum of the discrete alphabet average mutual information
over all possible discretizations or quantizations of $X^N$ and $Y^N$. 
If the joint distribution of $X^N$ and $Y^N$ is $\pi^N$, then we also write
$I(\pi^N )$ for $I(X^N,Y^N)$.

%The mutual information rate (if it exists) is
%then $I(X;Y)=\lim_{N\rightarrow \infty} N^{-1}I(X^N,Y^N)$. 
%%Not true without assumptions. But is it needed? E.g., is the process definition
%%of RDFs needed?

The Shannon rate-distortion 
function~\cite{Shannon:59} is defined for a stationary source $X$ by
%\begin{eqnarray}
%\lefteqn{
$$R_X(D)=%}\nonumber \\&=&
\infimum_N
N^{-1}R_{X^N} (D)=\lim_{N\rightarrow \infty} N^{-1}R_{X^N}(D)$$%\nonumber
%\\\lefteqn{
\begin{multline}R_{X^N}(D)%}\nonumber \\&=&
=
\infimum_{\pi^N:\pi^N\in \mathcal{P}(\mu_{X^N}), 
N^{-1}d(\pi^N)\le D}%Ed_N(\pi^N)\le D}
N^{-1}I(\pi^N) \label{rn}
\end{multline}%eqnarray}
where $\mathcal{P}(\mu_{X^N})$ is the collection of 
all joint distributions $\pi^N$ for $X^N,Y^N$ 
with first marginal distribution $\mu_{X^N}$.
\iffalse
where the infimum is over all joint distributions $\pi^N$ for $X^N,Y^N$ 
with first marginal distribution $\mu_{X^N}$ ($\pi^N\in \mathcal{P}(\mu_{X^N})$)
and
$N^{-1}E[d(X^N,Y^N)]\le D$. 
\fi
The dual  distortion-rate function is
%\begin{eqnarray*}
$$D_X(R)=%&=&
\infimum_N
N^{-1}D_{X^N} (R)=\lim_{N\rightarrow \infty} N^{-1}D_{X^N}(R)$$%\\
\begin{multline*}
D_{X^N}(R)=%\\%&=&
\infimum_{\pi^N:\pi^N\in \mathcal{P}(\mu_{X^N}), N^{-1}I(\pi^N)\le R}
N^{-1} Ed(\pi^N).%Ed_N(\pi^N).%Ed_N(X^N,Y^N).
\end{multline*}
Source coding theorems show that under suitable conditions
$\delta_X(R)=D_X(R)$. (See, e.g., 
\cite{GrayNO:75,Gray:75,Gray:90}
for source
coding theorems for stationary codes.)

%% The distortion-rate function is the more natural for the study of
%% systems which fix the rate (e.g., by fixing the encoder output
%% alphabet/decoder input alphabet) and minimize average distortion, but
%% the theory and algorithms for evaluation usually deal with the
%% rate-distortion function. Both are used here.

%\subsection{Evaluation of rate-distortion functions}
Csisz\'{a}r~\cite{Csiszar:74} provided quite general versions of
Gallager's~\cite{Gallager:68}
Kuhn-Tucker optimization for evaluating the rate-distortion functions
for finite dimensional vectors,
in particular restating  the optimization over
joint distributions $\pi^N$ as an optimization over the reproduction distribution
$\mu_Y^N$. When an optimizing reproduction distribution exists, it will be referred to as the \emph{Shannon optimal reproduction distribution}. Csisz\'{a}r provides conditions under which an optimizing distribution exists.

The following lemma and corollary are implied by
the proof of Csisz\'{a}r's Theorem 2.2 and
the extension of the reproduction space from compact metric to
Euclidean spaces discussed at the bottom of p. 66 of
\cite{Csiszar:74}. 
The lemma shows that if the distortion
measure  is a power of a metric derived from a norm, then there exists an
optimizing joint distribution
and hence also a
Shannon optimal reproduction distribution.  
In the corollary, the roles of distortion and mutual
information are interchanged to obtain the distortion-rate version of
the result.

\begin{lemma} \label{lemma:csiszar}
Let   $X$ be a random vector with an alphabet $A$ 
which is a finite-dimensional Euclidean space with
norm $\| x\|$. Assume the  reproduction 
alphabet $\hat{A}=A$ and a distortion measure
$d(x,y)=\|x-y\|^r$,
$r> 0$, such that $E[\|X\|^r]<\infty$. Then for any $D>0$ there exists a distribution $\pi$ on  
$A\times A$ achieving the
the minimum of (\ref{rn}). %(\ref{eq:rd1}). 
Hence for any $N$, a Shannon  $N$-dimensional optimal reproduction
distribution exists for the $N$th order rate-distortion function.
\end{lemma}

\begin{corollary}\label{corollary:tight}
Given the assumptions of the lemma, suppose that
$\pi^{(n)}$,  $n=1,2,\ldots$ is sequence of distributions on $A\times \hat{A}$ with marginals
$\mu_{X}$ and $\mu_{Y^{(n)}}$ for which for $n=1,2,\ldots$
\begin{eqnarray}
I(\pi^{(n)})&=&I(X,Y^{(n)}) \le  R, \label{Iconv}\\
\lim_{n\rightarrow \infty}E[d(X,Y^{(n)})]&=& D_{X}(R).\label{Dconv}
\end{eqnarray}
Then %$\pi^{(n)}$ and 
$\mu_{Y^{(n)}}$ has a subsequence that 
converges weakly
to a Shannon optimal reproduction distribution. If the Shannon distribution is unique, then
$\mu_{Y^{(n)}}$ converges weakly to it.
\end{corollary}

%% The result is proved by showing that the stated conditions imply that
%% any sequence of distributions $\pi^{(n)}$ has a weakly converging
%% subsequence and that the limiting distribution inherits the properties
%% of the individual $\pi^{(n)}$. If the Shannon optimal distribution is
%% unique, then we can assume that $\mu_{Y_0}^{(n)}$ converges weakly to
%% it.

%Note that if there is a unique Shannon optimal reproduction distribution, then
%any sequence of $\pi^{(n)}$ for which (\ref{Iconv}--\ref{Dconv}) hold must converge
%weakly to the optimal distribution.

\subsection{IID sources}
%\emph{IID sources}

If the process $X$ is IID, then
\begin{equation}\label{eq:rd1}
R_X(D)=R_{X_0}(D)=
\infimum_{\pi :\pi \in \mathcal{P}(\mu_{X_0}), 
Ed(X_0,Y_0)\le D}
I(X_0,Y_0).
\end{equation}
%Computation of the Shannon rate-distortion is significantly simplified in this case.
If a Shannon optimal distribution exists for the first-order rate distortion-function, then this
guarantees that it exists for all finite-order rate-distortion functions and that the optimal
$N$-th order distribution is simply the product distribution of $N$ copies of the first-order
optimal distribution. 

Rose~\cite{Rose:94} 
proved that for a continuous input random variable
and the squared error distortion, the Shannon optimal reproduction
distribution will be (absolutely) continuous only in the
special case where the Shannon lower bound to the rate distortion
function holds with equality, e.g., in the case of a Gaussian source
and squared error distortion. In other cases, the optimum reproduction
distribution is discrete, and for source distributions with bounded
support (e.g., the uniform $[0,1)$ source), the Shannon optimal
reproduction distribution will have finite support, that is, it will
be describable by a probability mass function (PMF) with a finite
domain. This last  result is originally due to Fix \cite{Fix:78}.
Rose proposed an algorithm using a form of annealing
which attempts to find the optimal finite alphabet directly by
operating on the source distribution,  avoiding the
indirect path of first discretizing the input distribution and then
performing a discrete Blahut algorithm --- the approach inherent
to the constrained alphabet rate-distortion theory and code design
algorithm of Finamore and Pearlman~\cite{FinamoreP:78}.  There is no
proof that Rose's annealing algorithm actually converges to the
optimal solution, but our numerical results support his arguments.

\section{Necessary Conditions for Optimal and Asymptotically Optimal
  Codes}
\label{sect:necessary}

A sliding-block code $(f,g)$ for source coding  is said to be optimum
if it yields an average distortion equal 
to the operational distortion-rate function, $D(f,g)= \delta_X(R)$. Unlike the simple scalar quantizer case
(or the nonstationary vector quantizer case), however, there are no simple conditions for guaranteeing
the existence of an optimal code. Hence usually it is of greater interest to consider codes that
are asymptotically optimal in the sense that their performance approaches the optimal in the limit,
but there might not be a code which actually achieves the limit. More precisely,
a sequence of rate-$R$ sliding-block codes 
$f_n,g_n$, $n=1,2,\ldots$,
for source coding 
is \emph{asymptotically optimal} (a.o.) if 
\begin{equation}
\lim_{n\rightarrow \infty}D(f_n,g_n)= \delta_X(R)=D_X(R)\label{eq:scao}.
\end{equation}
An optimal code (when it exists) is trivially asymptotically optimal and hence
any necessary condition for an asymptotically optimal sequence of codes also applies to
a fixed code that is optimal by simply equating every code in the sequence to the fixed code.

Similarly, a simulation code $g$ is optimal if
$\dbar (\mu_X , \mu_{\bar{g}(Z)})=%\Delta (X|Z)$
\Delta_{X\mid Z}(R)$
 and 
a sequence of codes $g_n$ is
asymptotically optimal if
\begin{equation}
\lim_{n\rightarrow \infty} \dbar (\mu_X , \mu_{\bar{g}_n(Z)})=%\Delta (X|Z).
\Delta_{X\mid Z}(R).
\end{equation}

In this section we exclusively focus on the squared error distortion
and assume that the real-valued stationary and ergodic process
$X=\{X_n\}$ has  finite variance.

\subsection{Process approximation}
The following lemma provides necessary conditions for asymptotically optimal 
codes which
are a slight generalization and elaboration of Theorem 1 of
Gray and Linder
\cite{GrayL:09}. A proof is provided in the Appendix.
\begin{lemma}\label{lemma:GL}(Condition 1)
Given a real-valued stationary ergodic 
process $X$, suppose that 
$f_n,g_n$
$n=1,2,\ldots$ 
is an asymptotically optimal sequence of stationary source codes
for  $X$
with encoder output/decoder input alphabet
$B$ of size $\|B\|=2^R$.
Denote  the resulting
reproduction processes by $\hat{X}^{(n)}$
and the $B$-ary  encoder output/decoder input processes by $U^{(n)}$. 
If $D_X(R)>0$, then 
\begin{eqnarray*}
\lim_{n\rightarrow \infty}  \dbar (\mu_X,\mu_{\hat{X}^{(n)}})&=&D_X(R)\\
\lim_{n\rightarrow \infty} H(\hat{X}^{(n)})
&=&
\lim_{n\rightarrow \infty} H(U^{(n)})=R\\
\lim_{n\rightarrow \infty}
\bar{d}_0 (U^{(n)},Z)&=&0,
\end{eqnarray*}
where $Z$ is an IID equiprobable process with alphabet size $2^R$.
\end{lemma}

These properties are quite intuitive:
\begin{itemize}
\item The process distance between a source and an approximately optimal reproduction of entropy rate less
than $R$ is close to the  Shannon distortion rate function.
Thus frequency-typical sequences of the reproduction should be as close as possible
to frequency-typical source sequences.
\item The entropy rate of an approximately optimal reproduction 
and of the resulting encoded $B$-ary process must be near the maximum possible value.
\item The sequence of encoder output processes approaches an IID equiprobable source
in the Ornstein process distance.  If $R=1$,  the encoder output bits should look like fair coin flips.

\end{itemize}

If $X$ is a $B$-process, then a sequence of a.o.\ simulation codes 
$g_n$ yielding a reproduction processes $\tilde{X}^{(n)}$ satisfies
$\lim_{n\rightarrow \infty}  \dbar (\mu_X,\mu_{\tilde{X}^{(n)}})=\Delta_{X|Z}(R)=D_X(R)$
and a similar argument to the proof of the previous lemma implies that
$
\lim_{n\rightarrow \infty} H(\hat{X}^{(n)})
=
H(Z)=R$.

\subsection{Moment conditions}
The next set of necessary conditions concerns the squared error
distortion and resembles a standard result for scalar and vector
quantizers (see, e.g., \cite{GershoG:92}, Lemmas 6.2.2 and
11.2.2). The proof differs, however, in that in the quantization case
the centroid property is used, while here simple ideas from linear
prediction theory accomplish a similar goal. Define in the usual way
the covariance $\mbox{COV}(X,Y)=E[(X-E(X))(Y-E(Y))]$.
%For this lemma we note the property for optimal codes to ease comparison with
%traditional results for scalar and vector quantization.
\begin{lemma}\label{lemma:moments}(Condition 2)
\iffalse
Given a real-valued stationary ergodic 
process $X$,
suppose that $(f,g)$ is an optimal 
 (with respect to squared error)
 stationary source code for a source $X$
which yields a reproduction $\hat{X}$ with entropy rate $H(\hat{X})\le R$. 
Then
\begin{eqnarray}
E(\hat{X}_0)&=&E(X_0) \label{eq:mean} \\ 
\mbox{COV}(X_0,\hat{X}_0)&=&\sigma_{\hat{X}_0}^2\\
%E(\hat{X}_0^2)&=&E(X^2)-\delta_X(R)\\
\sigma_{\hat{X}_0}^2&=&\sigma_{X_0}^2-D_X(R) 
\label{eq:variance}\label{eq:variance}
\end{eqnarray}
\fi
Given a real-valued stationary ergodic 
process $X$,
suppose that 
If $f_n,g_n$
is an asymptotically optimal sequence of codes
(with respect to squared error) yielding reproduction
processes $\hat{X}^{(n)}$
with entropy rate $H(\hat{X})\le R$, 
then
\begin{eqnarray}
\lim_{n\rightarrow \infty} E(\hat{X}_0^{(n)})&=&E(X_0)\label{eqlemma1} \\
\lim_{n\rightarrow \infty} 
{\mbox{COV} (X_0,\hat{X}_0^{(n)})
\over
\sigma_{\hat{X}_0^{(n)}}^2}&=&1\label{eqlemma2}\\
%\lim_{n\rightarrow \infty}  E((\hat{X}_0{(n)})^2)&=&E(X^2)-\delta_X(R)\nonumber\\
%&&\\
\lim_{n\rightarrow \infty}  \sigma_{\hat{X}_0^{(n)}}^2&=&\sigma_{X_0}^2-D_X(R)
\label{eqlemma3}
\end{eqnarray}
Defining the error as  $\epsilon_0^{(n)}=
\hat{X}_0^{(n)}-X_0$, then the necessary conditions become
\begin{eqnarray}
\lim_{n\rightarrow \infty} E(\epsilon_0^{(n)})&=&0\label{error1}\\
\lim_{n\rightarrow \infty} 
E(
\epsilon_0^{(n)}\hat{X}_0^{(n)}))&=&0\label{error2}\\
\lim_{n\rightarrow \infty} 
 \sigma_{\epsilon_0^{(n)}}^2&=&D_X(R)\label{error3}.
\end{eqnarray}
The results are stated for time $k=0$, but stationarity ensures that they hold for 
all times $k$.
\end{lemma}

%\emph{Remark:} The results parallel the classic moment results for
%optimal scalar quantization \cite{GershoG:92}.

%% necessary conditions for a sequence of codes to be
%% asymptotically optimal are that in the limit
%% \begin{itemize}
%% \item the means of the source and the reproduction are equal,
%% \item the correlation of the source and the reproduction 
%% with the means removed is the second moment
%% of the reproduction (the  error is uncorrelated with the reproduction), and
%% \item the variance of the source equals the sum of the expected
%%   squared error and the reproduction variance. 
%% Since the expected
%%   squared error of an asymptotically optimal sequence of codes is the
%%   distortion rate function, the result shows that a necessary
%%   condition for asymptotic optimality is satisfied by scaling the
%%   decoder so that the reproduction variance equals the input variance
%%   minus the distortion rate function.
%% \end{itemize}

\medskip

\noindent\emph{Proof:}
%We only prove the asymptotically optimal results 
%since they are the more important and since
%they imply the optimal results. 
For any encoder/decoder pair $(f_n,g_n)$ yielding a reproduction process
$\hat{X}^{(n)}$
\begin{eqnarray*}
D(f_n,g_n)&\ge&
\infimum_{a, b\in \R } 
D(f_n,a g_n+b )\nonumber \\
&\ge &
D_X(R)=\infimum_{f,g} D(f,g)
\end{eqnarray*}
where the second inequality follows since scaling a sliding-block decoder by
a real constant and adding a real constant
results in another sliding-block decoder with entropy rate no greater than
that of the input.
The minimization over $a$ and $b$
for each $n$
is solved  by standard linear prediction  techniques as
\begin{eqnarray}
a_n&=&{\mbox{COV} (X_0,\hat{X}_0^{(n)})
\over
\sigma_{\hat{X}_0^{(n)}}^2}\label{eq:andef}\\
b_n&=&
E(X_0)-a_nE(\hat{X}_0^{(n)})\label{eq:bdef},\\
\infimum_{a,b} 
D(f_n,a g_n+b)&=&
D(f_n,a_n g_n+b_n)\nonumber \\
&=&
\sigma_{X_0}^2-
a_n^2 \sigma_{\hat{X}_0^{(n)}}^2.
\end{eqnarray}
%Intuitively, we are simply choosing $a,b$ to yield the affine least squares estimate
%of the input $X_0$ given the reproduction $\hat{X}_0^{(n)}$. 
%An immediate property
%(which can be used to derive the optimal $a$) is the orthogonality property
%\begin{equation}\label{orthogonal}
%E[ (X_0-a_n\hat{X}_0^{(n)})\hat{X}_0^{(n)}]=0
%\end{equation}

Combining the above facts we have that
since $(f_n,g_n)$ is
an asymptotically optimal sequence, 
\begin{eqnarray}
D_X(R)&=&\lim_{n\rightarrow \infty} D(f_n,g_n)
%\nonumber \\&\ge&
\ge
\lim_{n\rightarrow \infty} D(f_n,a_n g_n+b_n)\nonumber\\
%&=&  \sigma_{X_0}^2- a_n^2\\
&\ge&
%\sigma_{X_0}^2-
D_X(R)\label{eq2}
\end{eqnarray}
and hence that both inequalities are actually equalities.
The final inequality (\ref{eq2}) being an equality yields
\begin{equation}\label{eq:a1}
\lim_{n\rightarrow \infty} 
a_n^2  \sigma_{\hat{X}_0^{(n)}}^2
=
\sigma_{X_0}^2-D_X(R).
\end{equation}
Application of asymptotic optimality and (\ref{eq:andef}) to
\begin{eqnarray*}
%\lefteqn{D(f_n,g_n)}\\
D(f_n,g_n)
&=&
E\left( (X_0- \hat{X}_0^{(n)})^2\right)\\
&=&
E\left( ([X_0- 
E(X_0)]
-[
\hat{X}_0^{(n)}-
E(\hat{X}_0^{(n)})] \right. \\
&&
\left.
\mbox{} +[E(X_0)-E(\hat{X}_0^{(n)})]
)^2\right)\\
&=&
\sigma_{X_0}^2+
 \sigma_{\hat{X}_0^{(n)}}^2-2\mbox{COV}(X_0,\hat{X}_0^{(n)})
 \\
 &&\mbox{} +
 [E(X_0)-E(\hat{X}_0^{(n)})]^2
 \end{eqnarray*}
 results in
 \begin{multline}
D_X(R)=\\
 \lim_{n\rightarrow \infty} \left(
 \sigma_{X_0}^2+ (1-2a_n)
 \sigma_{\hat{X}_0^{(n)}}^2
 + [E(X_0)-E(\hat{X}_0^{(n)})]^2\right). \label{eq:a2}
\end{multline}
 Subtracting (\ref{eq:a1}) from (\ref{eq:a2}) yields
 \begin{equation}
\lim_{n\rightarrow \infty} \left(
(1-a_n)^2 \sigma_{\hat{X}_0^{(n)}}^2
 + [E(X_0)-E(\hat{X}_0^{(n)})]^2\right)=0.\label{eq:a3}
\end{equation}
Since both terms in the limit are nonnegative, both must converge to zero since the sum does.
Convergence of the rightmost term in the sum proves (\ref{eqlemma1}).
Provided $D_X(R)<\sigma_{X_0}^2$, which is true if $R>0$,
(\ref{eq:a1}) and (\ref{eq:a3}) together imply that
$(a_n-1)^2/a_n^2$ converges to 0 and hence that 
\begin{equation}
\lim_{n\rightarrow \infty} a_n =
\lim_{n\rightarrow \infty} 
{\mbox{COV} (X_0,\hat{X}_0^{(n)})
\over
\sigma_{\hat{X}_0^{(n)}}^2}=1.\label{eq:a4}
\end{equation}
This proves (\ref{eqlemma2}) and with (\ref{eq:a2}) proves (\ref{eqlemma3}) and also that
\begin{equation}
\lim_{n\rightarrow \infty} 
\mbox{COV} (X_0,\hat{X}_0^{(n)})=\sigma_{X_0}^2-D_X(R).\label{eqlemma4}
\end{equation}

Finally consider the conditions in terms of the reproduction
error. Eq.\ (\ref{error1}) follows from
(\ref{eqlemma1}). Eq.\ (\ref{error2}) follows from
(\ref{eqlemma1})--(\ref{eqlemma4}) and some
algebra. Eq.\  (\ref{error3}) follows from (\ref{error1}) and the
asymptotic optimality of the codes.  \qed

If $X$ is a $B$-process so that
$\Delta_{X|Z}(R)=D_X(R)$, then a similar proof yields corresponding results for
the simulation problem. If  
$g_n$
is an asymptotically optimal (with respect to $\dbar_2$ distance)
sequence of stationary codes of an IID equiprobable source $Z$ with
alphabet $B$ of size $R=\log \| B \|$ which produce a simulated process
$\tilde{X}^{(n)}$, then
\begin{eqnarray*}
\lim_{n\rightarrow \infty} E(\tilde{X}_0^{(n)})
&=&
E(X_0)\\
\lim_{n\rightarrow \infty}
\sigma_{\tilde{X}_0^{(n)}}^2
&=&
\sigma_{X_0}^2-\Delta_{X|Z}(R).
\end{eqnarray*}
\iffalse
It is perhaps surprising that when finding the best matching process with constrained rate,
the second moments differ. 
\fi

\subsection{Finite-order distribution Shannon conditions for IID processes}
Several code design algorithms, including randomly populating a trellis to mimic the proof
of the trellis source encoding theorem~\cite{ViterbiO:74}, are based on the intuition that the guiding principle of
designing such a system for an IID source should be to produce a code with marginal reproduction
distribution close to a Shannon optimal reproduction distribution~\cite{WilsonL:77,FinamoreP:78,Pearlman:82}.
While highly intuitive, we are not aware of any rigorous demonstration to the effect that if a code is 
%optimal or 
asymptotically optimal, then necessarily its marginal reproduction distribution approaches that
of a  Shannon optimal. 
Pearlman~\cite{Pearlman:82} was the first to formally conjecture this
property of sliding-block codes.
The following result addresses this issue. It follows from standard inequalities 
and Csisz\'{a}r~\cite{Csiszar:74} as summarized in Corollary~\ref{corollary:tight}.
\begin{lemma}\label{lemma:aosc}(Condition 3a)
Given a real-valued IID process $X$ with distribution $\mu_X$, assume that
$f_n,g_n$
is an asymptotically optimal 
sequence of stationary source encoder/decoder pairs  with common
encoder output/decoder input alphabet $B$ of size $R=\log \| B \|$ which produce a reproduction process
$\hat{X}^{(n)}$.
Then a subsequence of the marginal distribution of the reproduction process,
$\mu_{\hat{X}_0^{(n)}}$
converges weakly  and in $\mathcal{T}_2$
to a Shannon optimal reproduction distribution. 
If the Shannon optimal reproduction distribution is unique, then 
$\mu_{\hat{X}_0^{(n)}}$ converges to it.

\end{lemma}
\noindent\emph{Proof:}
Given the asymptotically optimal sequence of codes, let $\pi_n$ 
denote the induced process joint distributions on $(X,\hat{X}^{(n)})$.
The encoded process has alphabet size $2^R$ and hence entropy rate
less than or equal to 
$R$. Since coding cannot increase entropy rate, the entropy rate of the reproduction (decoded)
process is also less than or equal to $R$. By standard information theoretic inequalities 
(e.g., \cite{Gray:90}, p. 193), since the input process is IID we have for all $N$ that
\begin{eqnarray}
{1\over N}I(\pi_n^N)
&=&
{1\over N}I(X^N,\hat{X}^N)\nonumber
\ge
{1\over N}\sum_{i=0}^{N-1}I(X_i,\hat{X}_i^{(n)}\nonumber
)\\
&=&
I(X_0,\hat{X}_0^{(n)})
=I(\pi_n^1).
\end{eqnarray}
The leftmost term converges to the mutual information rate between the input
and reproduction, which is bound above by the entropy rate of the output so that
$
I(X_0,\hat{X}_0^{(n)}) \le R, \mbox{ all } n.
$
Since the code sequence is asymptotically optimal, (\ref{eq:scao})
holds. Thus the sequence of joint distributions $\pi_{n}$ for
$(X_0,\hat{X}_0^{(n)})$ meets the conditions of
Corollary~\ref{corollary:tight} and hence $\mu_{\hat{X}_0^{(n)}}$
has a subsequence which
converges weakly to a Shannon optimal distribution.
If the Shannon optimal distribution $\mu_{Y_0}$ is unique, 
then every subsequence of
of $\mu_{\hat{X}_0^{(n)}}$ has a further subsequence which converges
to $\mu_{Y_0}$, which implies that $\mu_{\hat{X}_0^{(n)}}$ converges weakly
to $\mu_{Y_0}$.
The moment conditions %(\ref{eq:mean}), (\ref{eq:variance}), 
(\ref{eqlemma1}) and (\ref{eqlemma3})) of
Lemma~\ref{lemma:moments} imply that $E[(\hat{X}_0^{(n)})^2]$
converges to $E[(\hat{X}_0)^2]$.
The weak convergence of
a subsequence of $\mu_{\hat{X}^{(n)}}$ (or the sequence itself)
and the convergence of
the second moments imply convergence in $\mathcal{T}_2$~\cite{Villani:09}.
\qed

Since the source is IID, the $N$-fold product of a one-dimensional
Shannon optimal distribution is 
an $N$-dimensional Shannon optimal distribution. If the Shannon optimal marginal distribution
is unique, then so is the $N$-dimensional Shannon optimal distribution. Since 
Csisz\'{a}r's~\cite{Csiszar:74} results hold for the $N$-dimensional case, we immediately have
the first part of the following corollary.
\begin{corollary}\label{lemma:finiteorder}(Condition 3b)
Given the assumptions of the lemma, for any positive integer $N$ let
$\mu_{\hat{X}^{(n)}}$ denote the $N$-dimensional joint distribution of
the reproduction process $\hat{X}^{(n)}$. Then a subsequence of the
$N$-dimensional reproduction distribution $\mu_{\hat{X}^{(n)}}$
converges weakly and in $\mathcal{T}_2$ to the $N$-fold product of a Shannon optimal marginal
distribution (and hence to an $N$-dimensional Shannon optimal
distribution).  
If the one dimensional Shannon optimal
distribution is unique, then $\mu_{\hat{X}^{(n)}}$ converges weakly
and in $\mathcal{T}_2$ 
to its $N$-fold product distribution.
\end{corollary}
\noindent\emph{Proof:}
The moment conditions %(\ref{eq:mean}), (\ref{eq:variance}), 
(\ref{eqlemma1}) and (\ref{eqlemma3})) of
Lemma~\ref{lemma:moments} imply that $E[(\hat{X}_k^{(n)})^2]$
converges to $E[(\hat{X}_k)^2]$ for $k=0,1,\ldots , N-1$.
The weak convergence of the $N$-dimensional distribution 
of a subsequence of $\mu_{\hat{X}^{(n)}}$ (or the sequence itself)
and the convergence of
the second moments imply convergence in $\mathcal{T}_2$
\cite{Villani:09}.\qed 

There is no counterpart of this result for optimal codes as opposed to asymptotically optimal
codes. Consider the Gaussian case where the Shannon optimal distribution is a product
Gaussian distribution with variance $\sigma_X^2-D_X(R)$.
If a code were optimal, then for each $N$ the resulting $N$th order
reproduction distribution would  
have to
equal the Shannon
product distribution. But if this were true for all $N$, the reproduction would
have to be the IID process with the Shannon marginals, but that process has infinite entropy rate.

If $X$ is a $B$-process, then a 
small variation on the proof yields similar results for the simulation problem:
given an IID target source $X$,
the $N$th order joint distributions $\mu_{\tilde{X}^{(n)}}$ of an asymptotically optimal sequence of 
constrained rate simulations $\tilde{X}^{(n)}$ will have a subsequence that converges
weakly and in $\mathcal{T}_2$ to an $N$-dimensional Shannon optimal
distribution. 

%% If the Shannon optimal marginal distribution is unique, then
%% $\mu_{\tilde{X}^{(n)}}$ will converge weakly to the Shannon optimal
%% distribution.  Lemma~\ref{lemma:white2} in the Appendix shows that if a
%% sequence of $N$-dimensional distributions for a random vector
%% converges in $\mathcal{T}_2$ to an IID distribution with finite-second
%% moments, then all covariances except for zero lag converge to 0. This
%% leads to the next result.

\subsection{Asymptotic uncorrelation}
The following theorem proves a result that   has often been
assumed or claimed to be a property of optimal codes.
Define as usual the covariance function of the stationary
process $\hat{X}^{(n)}$ by
$
K_{\hat{X}^{(n)}}(k)
=
\mbox{COV} (\hat{X}_i^{(n)},
\hat{X}_{i-k}^{(n)})
$ for all integer $k$.

\begin{lemma}\label{lemma:white}(Condition 4)
Given a real-valued IID process $X$ with distribution $\mu_X$, assume that
$f_n,g_n$
is an asymptotically optimal 
sequence of stationary source encoder/decoder pairs  with common
alphabet $B$ of size $R=\log \| B \|$ which produce a reproduction process
$\hat{X}^{(n)}$. 
For all $k\neq 0$,
\begin{equation}
\lim_{n\rightarrow \infty}
K_{\hat{X}^{(n)}}(k)
=
0
\end{equation}
and hence the reproduction processes are asymptotically uncorrelated. 
\end{lemma}

\noindent\emph{Proof.} If the Shannon optimal distribution is unique,
then $\mu_{\hat{X}^{(n)}}$ converges in $\mathcal{T}_2$ to the
$N$-fold product of the Shannon optimal marginal distribution by
Corollary~\ref{lemma:finiteorder}. As Lemma~\ref{lemma:white2}
in the Appendix shows, this implies the convergence of
$K_{\hat{X}^{(n)}}(k)=
\mbox{COV} (\hat{X}_k^{(n)},
\hat{X}_{0}^{(n)})$  to 0 for all $k\neq 0$.\qed

\smallskip

Taken together these necessary conditions provide straightforward tests for code construction algorithms.
Ideally, one would like to prove that a given code construction satisfies these properties,
but so far this has only proved possible for the Shannon optimal reproduction distribution property --- as exemplified in the next section. The remaining properties, however, can be easily demonstrated
numerically.% for the codes developed next. 

\section{An Algorithm for Sliding-Block Simulation and Source Decoder
  Design}
\label{sect:algorithm}

%In the previous section it was demonstrated that necessary conditions for good codes
%for stationary ergodic sources
%is that they yield a reproduction process with the same mean but a reduced variance as the original
%source and that they yield a binary channel process that is close to fair coin flips. For an IID source,
%it was also shown that the marginal distribution of the reproduction process must
%be close to that of the Shannon optimal distribution for the rate-distortion function.

We begin with a sliding-block simulation code which approximately satisfies
the Shannon marginal distribution
necessary condition for optimality. Matching the code with a Viterbi algorithm
(VA) encoder then yields a trellis source encoding system. 

\subsection{Sliding-block simulation code/source decoder}
Consider a sliding-block
code $g_L$ of length $L$
of an equiprobable binary IID process $Z$ which produces an output
process $\tilde{X}$ defined by
\begin{equation}
\tilde{X}_n=%g(Z_{n-L+1}, Z_{n-K+1},\cdots , Z_{n-1}, Z_n),
g( Z_n,Z_{n-1},\cdots , 
Z_{n-L+1}),
\end{equation}
where the notation makes sense even if $L$ is infinite, in which case $g$ views a semi-infinite
binary sequence. Since the processes are stationary, we emphasize the
case $n=0$.
Suppose that the ideal distribution for $\tilde{X}_0$ is given by a CDF $F$,
for example the CDF corresponding to the Shannon optimal marginal reproduction
distribution of Lemma~\ref{lemma:csiszar}.
Given a CDF $F$, define the (generalized) inverse CDF
$F^{-1}$ as $F^{-1}(u)=\inf\{r\colon F(r)\ge u\}$ for $0<u<1$.  
If $U$ is a uniformly distributed continuous
random variable on $(0,1)$, then  the
random variable $F^{-1}(U)$ has CDF $F$. The CDF can be approximated by considering the
binary $L$-tuple 
$u^L=(u_0,u_1,\ldots , u_{L-1})$ 
comprising the shift register entries as the binary expansion of a number in $(0,1)$:
\begin{equation}
b(u^L)=
\sum_{i=0}^{L-1}
u_i 2^{-i-1}
+
 2^{-L-1}
,
\end{equation}
and defining 
\begin{multline}
g( Z_n,Z_{n-1},\cdots , 
Z_{n-L+1})%Z_{n-L+1},\cdots , Z_{n-1}, Z_n)
=\\
F^{-1}(b( Z_n,Z_{n-1},\cdots , 
Z_{n-L+1})).
\end{multline}
If the $Z_n$ is a fair coin flip process, the discrete random variable 
$b( Z_n,Z_{n-1},\cdots , 
Z_{n-L+1})$
is uniformly distributed on the discrete set
$\{2^{-L-1},2^{-L-1}+2^{-L},2^{-L-1}+2\times 2^{-L},\cdots ,2^{-L-1}+ 1-2^{-L}\}$, 
that is, it is a discrete approximation
to a uniform $(0,1)$ that improves as $L$ grows, and the distribution of
$g( Z_n,Z_{n-1},\cdots , 
Z_{n-L+1})%Z_{n-L+1}, \cdots , Z_{n-1}, Z_n)
$ converges weakly to
$F$, satisfying a necessary condition for an asymptotically
optimal sequence of codes. If $L$ is infinite, then the marginal distribution
will correspond to the target distribution exactly!
This fulfills the necessary condition of weak convergence for an 
asymptotically optimal code 
of Lemma~\ref{lemma:aosc}.

%%move permutation discussion to here, put before trellis discussion
The code as described thus far only provides
the correct approximate marginals;  it does not provide joint distributions 
that match the Shannon optimal joint distribution --- nor can it exactly
since it cannot produce independent pairs. We adopt %??adapt 
a heuristic
aimed at making pairs of reproduction samples as independent as possible
%in the sense of $\dbar_2$ approximation 
by modifying the code in a way
that decorrelates successive reproductions and hence attempts to satisfy the
necessary condition of Lemma~\ref{lemma:white}.
Instead of applying the
inverse CDF directly to the binary shift register contents, we first
permute the binary vectors, that is, the codebook of all $2^L$
possible shift register contents is permuted by an invertible
one-to-one mapping $\mathcal{P}:\{0,1\}^L\rightarrow \{0,1\}^L$ and
the binary vector $\mathcal{P}(u^L)$ is used to generate the discrete
uniform distribution.  A randomly chosen permutation $\mathcal{P}$ is used, but
\emph{once chosen it is fixed} so that sliding-block decoder is truly
stationary.  Such a random choice to obtain a code that is then used for all time is analogous to the traditional Shannon block source coding proof of randomly choosing a decoder codebook which is then used for all time.
Thus our decoder is
\begin{multline}
g(Z_n,Z_{n-1},\cdots,Z_{n-L+1})=\\
F_{Y_0}^{-1}(b( \mathcal{P}(Z_n,Z_{n-1},\cdots,Z_{n-L+1})),
\end{multline}
where $F_{Y_0}(y)$ is a Shannon optimal reproduction distribution obtained either analytically (as in the Gaussian
case) or from the Rose algorithm (to find the optimum finite support).

Intuitively, the permutation should make the resulting sequence of arguments of the 
mapping (the number in $(0,1)$ constructed from the permuted binary symbols) resemble an independent sequence
and hence cause the sequence of branch labels to locally appear to be independent. 
The goal is to satisfy the necessary conditions on joint reproduction distributions of 
Corollary~\ref {lemma:finiteorder},
but we have no proof that the proposed construction has this property.
The experimental results to be described show excellent performance 
approaching the Shannon rate-distortion
bound and show that the branch labels are %at least 
indeed uncorrelated. %, as required by Lemma~\ref{lemma:white}.
The permutation is implemented easily by permuting the table entries defining $g$.
For the constrained-rate simulation problem, the permutation does not change the marginal distribution 
of the coder output, which still converges weakly to the Shannon optimal reproduction distortion as
$L\rightarrow \infty$, even in the Gaussian case. 
This approach is in the spirit of Rose's mapping approach
to  finding
the rate-distortion function~\cite{Rose:94}
since it involves discretizing a 
continuous uniform random variable which is the argument to a mapping 
into the reproduction space, rather than discretizing the
source. 

The decoder design involves no training (assuming that the Shannon optimal
marginal distribution is known). % and yields a stationary code. 

\subsection{Trellis encoding}
If the decoder of a source
coding system is a finite-length sliding-block code, then encoding can
be accomplished using a VA search of the trellis
diagram labeled by the available decoder outputs.
A trellis is a directed graph showing the action of a finite-state
machine with all but the newest %leftmost (newest) 
symbol in the shift
register constituting the state and the 
newest symbol %leftmost symbol in the figure
being the input. Branches connecting each state are labeled by the
output (or an index for the output in a reproduction codebook)
produced by receiving a specific input in a given
state.
As usually implemented, 
the VA
yields a block encoder matched to the sliding-block decoder.
A source coding system having this form is a
\emph{trellis source encoding system}. 

The theoretical properties of asymptotically optimal codes developed here are for 
the combination of stationary encoder and decoder, but our numerical results
use the traditional trellis source encoding structure of a block VA matched to a
sliding-block decoder. In fact, we perform a full search on the entire  test sequence since this provides the smallest possible average distortion encoding using the given decoder. This apparent mismatch of a theoretical emphasis on overall stationary codes with a hybrid stationary decoder/block encoder merits explanation. First, our emphasis is on decoder design and given a sliding-block decoder, no encoder can yield smaller average distortion than a matched VA algorithm operating on the entire dataset. Available computers permit such an implementation for datasets and decoders of interesting size. 
A source coding theorem for a block Viterbi encoder and a stationary decoder may be found  in \cite{Gray:77}.
Second,
using standard techniques for converting a block code into a sliding-block code,
a VA block encoder can be approximated as closely as desired by a sliding-block code. Such approximations originate in Ornstein's proof of his isomorphism theorem~\cite{Ornstein:73,Ornstein:75} and have been developed specifically for tree and trellis encoding systems, e.g., in Section VII of \cite{Gray:75}, 
and for block source codes in general in \cite{Shields:79,Gray:90}.
These constructions embed a good block code into a stationary structure by means of a punctuation sequence which inserts rare spacing between long blocks --- which in practice would mean adding significant computational complexity to the straightforward Viterbi search of the approximately optimal decoder output. Other, simpler,  means of stationarizing the VA such as incremental tree and trellis encoding \cite{AndersonB:75,Gray:77} have been considered,
but they are not supported by coding theorems. Experimentally, however, they have been shown to provide essentially the same performance as the usual block Viterbi encoder.
The hybrid code with a VA encoder and a stationary decoder remains the simplest implementation and takes full advantage of the stationary decoder which is designed here. Third, our necessary conditions for optimal stationary codes focus on the reproduction process and hence depend  on the decoder and its correspondence to an optimal simulation code. 
The Associate Editor has pointed out that the theoretical results for stationary codes can likely be reconciled with our use of a block encoder/stationary decoder by extending our necessary conditions to incorporate hybrid codes such as fixed-rate (or variable-rate \cite{YangZ:99}) trellis encoding systems by replacing our marginal distributions by average marginal distributions. We suspect this is true and that our results will hold for any coding structure yielding asymptotically mean stationary processes,
but we have chosen not to attempt this here in the interests of simplicity and clarity.  

A brief overview
of the history of trellis source encoding provides useful context for comparing
the numerical results. A stationary decoder produces a time-invariant
trellis and
trellis branch labels  that do not change with time.  
The original
1974 source coding theorem for trellis encoded IID
sources~\cite{ViterbiO:74} was proved for 
\emph{time-varying codes} by using
a variation of Shannon random coding --- successive levels of the trellis
were labeled randomly based on IID random variables chosen according
to the test channel output distribution arising in the evaluation of
the Shannon rate-distortion function.

Early research on trellis
encoding design
was concerned with time-varying trellises, reflecting the structure of the coding
theorem. 
In particular, Wilson and Lytle~\cite{WilsonL:77} %Wilson:75,}
populated their trellis using IID random labels chosen according to
the Shannon optimal reproduction distribution.  A later source coding
theorem for time-invariant trellis encoding \cite{Gray:77} was based
on the sliding-block source coding theorem \cite{GrayNO:75,Gray:75}
and was purely an existence proof; it did not suggest any
implementable design techniques. Two early techniques for
time-invariant code design were the fake process
design~\cite{LindeG:78} and a Lloyd clustering approach conditioned on
the shift register states~\cite{Stewart:81,StewartGL:82}.  The former
technique was based on a heuristic argument involving optimal
simulation and the $\dbar$-distance formulation of the operational
distortion rate function. The idea was to color a trellis with a
process as close in $\dbar$ as possible to the original source. While
the goal is correct, the heuristic adopted to accomplish it was
flawed: the design attempted to match the marginal distribution and
the power spectral density of the reproduction with those of the
original source. As pointed out by Pearlman~\cite{Pearlman:82} and
proved in this paper, the marginal distribution of the trellis labels
should instead match the Shannon optimal distribution, not the
original source distribution.

%%--Redundant, already said
%Pearlman demonstrated
%experimentally and with supporting theory that better performance resulted from coloring
%he trellis so as to yield a marginal reproduction distribution close to the Shannon optimal
%reproduction distribution, rather than to the original source density~\cite{Pearlman:82}. 
Pearlman's
theoretical development~\cite{Pearlman:82} was based on his and 
Finamore's constrained-output alphabet rate-distortion
~\cite{FinamoreP:78}, which involved a prequantization step prior to to designing
a trellis encoder for the resulting finite-alphabet process. 
Pearlman provided a coding theorem and an implementation for a time-invariant trellis encoding,
but used the artifice of a subtractive dithering sequence to ensure the necessary independence
of successive trellis branch labels over the code ensemble. Because of the dithering,
the overall code is not time-invariant.

Marcellin and Fisher in 1990~\cite{MarcellinF:90} introduced 
%a newdesign philosophy, 
trellis-coded quantization (TCQ) %, for trellis encoding 
based on an analogy with coded modulation in the dual problem
of trellis decoding for noisy channels. The technique provided a coding technique of much
reduced complexity that has since become one of the most popular
compression systems for a variety of signals. The dual code argument
is strong, however, only for the uniform case, but variations of the idea have
proved quite effective in a variety of systems.  TCQ has a default
assignment of reproduction values to trellis branches using a
Lloyd-optimized quantizer, but the levels can also be optimized. 
%Both default and optimized levels are considered here.

%% We include TCQ in our comparisons, but our emphasis is on trying to
%% optimize performance for a known source distribution rather than on
%% low complexity implementation.
%% %The notion of ``reasonable'' complexity has grown
%% %with hardware and software improvements, so it is of interest to see how nearly optimal performance
%% %can be when all the computing power on a typical modern machine is used. 

Some techniques,
including TCQ in our experiments, tend to reach a performance ``plateau'' in that performance improvement with complexity becomes negligible well before the complexity becomes burdensome. 
In TCQ this can be attributed to constraints placed on the system to ensure low complexity.
The technique introduced here has not (yet) shown any such plateau.

More recently, van der Vleuten and Weber~\cite{vanderVleutenW:95}
combined the fake process intuition with TCQ to obtain improved
trellis coding systems for IID sources. They incorrectly stated that
\cite{LindeG:78} had shown that a necessary condition for optimality
for trellis reproduction labels for coding an IID source is that the
reproduction process be uncorrelated (white) when the branch labels
are chosen in an equiprobable independent fashion.  This is indeed an
intuitively desirable property and it was used as a guideline in
\cite{LindeG:78} --- but it was not shown to be necessary.  Eriksson et
al.~\cite{ErikssonAG:07} used linear congruential (LC) recursions to
generate trellis labels %(effectively indexes of reproduction values)
and reproduction values to develop the best codes of the time for IID
sources to date by establishing a set of ``axioms'' of desirable
properties for good codes (including a flat reproduction spectrum) and
then showing that a trellis decoder based on an inverse CDF of a
sequence produced by linear recursion relations meets the conditions.
Because of the CDF matching and spectral control, the system can also
be viewed as a variation on the fake process approach. Eriksson et
al.\ observe that a problem with TCQ is the constrained ability to
increase alphabet size for a fixed rate and they argue that larger
alphabet size can always help. This is not correct in general,
although it is for the Gaussian source where the Shannon optimal
distribution is continuous. For other sources, such as the uniform,
the Shannon optimal has finite support and optimizing for an alphabet
that is too large or not the correct one will hurt in general.  As with TCQ, the
approach allowed optimization of the reproduction values assigned to
trellis branch labels.

\section{Numerical Examples}
\label{sect:numerical}
The random permutation trellis encoder was designed for three common IID test sources:
Gaussian, uniform, and Laplacian. The results 
in terms of both mean squared error (MSE) and signal-to-noise ratio (SNR) 
are reported for various  shift register
lengths $L$ indicated by RP\_$L$, here RP\_$L$ stands for random permutation
trellis coding algorithm with shift register length $L$.
The test sequences were all of length $10^6$. The results for Gaussian, uniform 
and Laplacian sources are shown in Table~\ref{table:gauss}, ~\ref{table:uniform}
 and ~\ref{table:laplace} respectively.

Each test result is from one random permutation; repeating the test
with different random permutations has produced almost identical
results.  E.g.\ for IID Gaussian source, $R=1$, $L=16$, a total of
$20$ test runs have returned MSE in the range between $0.2629$ and
$0.2643$, with an average of $0.2634$.

\iffalse
This is due to the long test sequence and the fact that at any reasonable
shift register length,
the number of possible permutation is extremely large. E.g. for $L=8$,
there are $2^8!$ possible permutations. So a randomly chosen permutation
has probability close to 1 to be a typical permutation.
\fi

The distortion-rate function $D_X(R)$ for all three sources are also listed in 
the tables. For uniform and Laplacian sources, $D_X(R)$ are numerical 
estimations produced by the Rose algorithm, in both cases, the reported 
distortions are slightly lower in comparison to the results reported 
in \cite{MarcellinF:90,NollZ:78} calculated using the 
Blahut algorithm~\cite{Blahut:72}.

The rate $R=1$ results of the random permutation trellis coder are compared to previous results of the 
linear congruential trellis codes (LC) of Eriksson, Anderson, and Goertz~\cite{ErikssonAG:07},
trellis coded quantization (TCQ) by
Marcellin and Fischer~\cite{MarcellinF:90}, trellis source encoding by Pearlman~\cite{Pearlman:82} based
on constrained reproduction alphabets and matching the Shannon optimal marginal distribution, 
a Lloyd-style clustering algorithm conditioned on trellis states by Stewart et al.~\cite{Stewart:81,StewartGL:82},
and the
Linde-Gray fake process design~\cite{LindeG:78}.
The rate $R=2$ results are compared with Eriksson et al.'s  LC codes and
Marcellin and Fisher's TCQ.
The rate $R=3,4$ results are compared with TCQ which are the only available previous results for these rates.

Eriksson et al.'s LC codes use 512 states for $R=1$ and 256 states for $R=2$, 
which are equivalent to shift register length $L=10$ in both cases. 
Marcellin's TCQ uses 256 states for all rates, corresponding to shift register 
length 9,10,11,12 for rate 1,2,3,4 respectively. Pearlman's results  and Stewart's results are for $L=10$, and 
Linde/Gray uses a shift register of length~9.
The shift register length $L$ is indicated as a subscript for all results.

In the Gaussian example, there are $2^{L}$ reproduction levels in the
random permutation codes, the result of taking the inverse Shannon
optimal CDF, that of a Gaussian zero mean random variable with
variance $1-D_X(R)$, and evaluating it at $2^L$ uniformly spaced
numbers in the unit interval. For the uniform source, there are
3, 6, 12, and 
24 reproduction points for rates 1, 2, 3, 4 bits chosen by the Rose
algorithm for evaluating the first order rate-distortion
function. Similarly, for the Laplacian source , there are 9, 17, 31,
and 55 reproduction points for rates 1,2,3,4 bits, respectively.  For
rates $R=2,3,4$ bits, the trellis has $2^R$ outgoing branches from
each node and $2^R$ incoming branches to each node. $R$ new bits are
shifted into the shift register and $R$ old bits are shifted out at
each transition. The Viterbi search now merges $2^R$ paths at each
node compared to just $2$ paths in the $1$ bit case. The number of
states in the trellis is $2^{(L-R)}$, for the trellis structures with
the same number of states; the $R=2$ trellis has shift register length
1 bit longer compared to the $R=1$ trellis and also has twice the
number of branches/reproduction levels.

Eriksson et al.'s LC codes use $2^{L-1}$ reproduction points, the Linde/Gray fake process design uses $2^L$ reproduction points, 
in both cases, the reproduction points are generated by taking the inverse CDF of the source,
evaluating it in the unit interval, and then multiplying with a scaling factor. 
Stewart also uses $2^L$ reproduction points, but the reproduction points are 
obtained through an iterative Lloyd-style training algorithm. Pearlman uses 
a simpler 4 symbol reproduction alphabet, produced by the Blahut
algorithm. Marcellin's TCQ uses $2^{R+1}$ reconstruction symbols,
which are the outputs  
of the Llyod-Max quantizer. In both LC codes and TCQ, numerical optimization 
of the reproductions values were used to improve the results. The optimized 
results for LC codes and TCQ are listed in the tables with the notation ``(opt)''.

The TCQ\_9 and TCQ(opt)\_9 results are from  Marcellin and Fisher's TCQ
paper~\cite{MarcellinF:90}. 
The TCQ results at shift register length 12,16,20,24 are asterisked since
they are from our own implementation of the TCQ following descriptions
in~\cite{MarcellinF:90}. In our implementation, the default reproduction
values, not the optimized ones were used. The TCQ results are clearly
showing a performance "plateau" as the shift register length increases. 

\begin{table}
\begin{center}
\begin{tabular}{|p{1.5cm}|c|p{1.15cm}|p{1.15cm}|}
\hline
           & Rate(bits) & MSE    & SNR(dB) \\
\hline
\hline
RP\_8    &  1  &  0.2989  &  5.24  \\
\hline
RP\_9    &  1  &  0.2913  &  5.36  \\
\hline
RP\_10    &  1  &  0.2835  &  5.47  \\
\hline
RP\_12    &  1  &  0.2740  &  5.62  \\
\hline
RP\_16    &  1  &  0.2638  &  5.79  \\
\hline
RP\_20    &  1  &  0.2582  &  5.88  \\
\hline
RP\_24    &  1  &  0.2557  &  5.92  \\
\hline
RP\_28    &  1  &  0.2542  &  5.95  \\
\hline
$D_X(R)$   & 1   & 0.25    &  6.02    \\
\hline
TCQ\_9        & 1   & 0.3105  &  5.08      \\
\hline
TCQ(opt)\_9   & 1   & 0.2780  &  5.56      \\
\hline
TCQ\_$12^{\ast}$        & 1   & 0.3088  & 5.10       \\
\hline
TCQ\_$16^{\ast}$        & 1   & 0.3072  & 5.13       \\
\hline
TCQ\_$20^{\ast}$        & 1   & 0.3064  & 5.14       \\
\hline
TCQ\_$24^{\ast}$        & 1   & 0.3060  & 5.14       \\
\hline
Pearlman\_10   & 1   & 0.292   &  5.35      \\
\hline
Stewart\_10    & 1   & 0.293   &  5.33      \\
\hline
Linde/Gray\_9 & 1   & 0.31    &  5.09      \\
\hline
LC\_10         & 1   & 0.2698  &  5.69      \\
\hline
LC(opt)\_10     & 1   & 0.2673  &  5.73      \\
\hline
\hline
           & Rate(bits) & MSE    & SNR(dB) \\
\hline
\hline
RP\_10    &  2  &  0.0797  &  10.98  \\
\hline
RP\_24    &  2  &  0.0646  &  11.90  \\
\hline
$D_X(R)$   & 2   & 0.0625    &  12.04    \\
\hline
TCQ\_10        & 2  &  0.0873  &  10.59      \\
\hline
TCQ(opt)\_10   & 2  &  0.0787  &  11.04      \\
\hline
LC(opt)\_10     & 2  &  0.0690  &  11.61      \\
\hline
\hline
RP\_11    &  3  &  0.0208  &  16.81  \\
\hline
RP\_24    &  3  &  0.0162  &  17.90  \\
\hline
$D_X(R)$   & 3   & 0.0156    &  18.06    \\
\hline
TCQ\_11        & 3  & 0.0237   &  16.25      \\
\hline
TCQ(opt)\_11   & 3  & 0.0217   &  16.64      \\
\hline
\hline
RP\_12    &  4  &  0.0054  &  22.71  \\
\hline
RP\_24    &  4  &  0.0041  &  23.92  \\
\hline
$D_X(R)$   & 4   & 0.0039    &  24.08    \\
\hline
TCQ\_12        & 4  & 0.0062   &  22.05      \\
\hline
\end{tabular}
\end{center}
\caption{Gaussian Example}\label{table:gauss}
\vspace{-0.2in}
\end{table}

\begin{table}
\begin{center}
\begin{tabular}{|p{1.5cm}|c|p{1.15cm}|p{1.15cm}|}
\hline
           & Rate(bits) & MSE    & SNR(dB) \\
\hline
\hline
RP\_8    &  1  &  0.0203  &  6.13  \\
\hline
RP\_9    &  1  &  0.0195  &  6.30  \\
\hline
RP\_10    &  1  &  0.0190  &  6.42  \\
\hline
RP\_12    &  1  &  0.0184  &  6.55  \\
\hline
RP\_16    &  1  &  0.0179  &  6.69  \\
\hline
RP\_20    &  1  &  0.0176  &  6.75  \\
\hline
RP\_24    &  1  &  0.0175  &  6.78  \\
\hline
RP\_28    &  1  &  0.0174  &  6.79  \\
\hline
$D_X(R)$   & 1   & 0.0173  &  6.84  \\
\hline
TCQ\_9        & 1   & 0.0194    &  6.33       \\
\hline
TCQ(opt)\_9   & 1   & 0.0183    &  6.58       \\
\hline
LC\_10        & 1   & 0.0191    &  6.40       \\
\hline
LC(opt)\_10    & 1   & 0.0179    &  6.67       \\
\hline
\hline
           & Rate(bits) & MSE    & SNR(dB) \\
\hline
\hline
RP\_24    &  2  &  4.02e-03  &  13.17  \\
\hline
$D_X(R)$   & 2   & 3.96e-03  &  13.23  \\
\hline
TCQ\_10        & 2   &  4.24e-03 &  12.93      \\
\hline
TCQ(opt)\_10   & 2   &  4.18e-03 &  13.00      \\
\hline
LC(opt)\_10    & 2   &  4.13e-03 &  13.05      \\
\hline
\hline
RP\_24    &  3  &  9.70e-04  &  19.34  \\
\hline
$D_X(R)$   & 3   & 9.46e-04  &  19.45  \\
\hline
TCQ\_11        & 3   & 10.0e-04 &  19.20       \\
\hline
TCQ(opt)\_11   & 3   & 9.95e-04 &  19.23       \\
\hline
\hline
RP\_24    &  4  &  2.39e-04   &  25.43  \\
\hline
$D_X(R)$   & 4   & 2.35e-04  &  25.50  \\
\hline
TCQ\_12   & 4   & 2.44e-04  &  25.34    \\
\hline
\end{tabular}
\end{center}
\caption{Uniform $[0,1)$ Example}\label{table:uniform}
\vspace{-0.2in}
\end{table}

\begin{table}
\begin{center}
\begin{tabular}{|p{1.5cm}|c|p{1.15cm}|p{1.15cm}|}
\hline
           & Rate(bits) & MSE    & SNR(dB) \\
\hline
\hline
RP\_8    &  1  &  0.2946  &  5.31  \\
\hline
RP\_9    &  1  &  0.2789  &  5.55  \\
\hline
RP\_10    &  1  &  0.2671  &  5.73  \\
\hline
RP\_12    &  1  &  0.2532  &  5.97  \\
\hline
RP\_16    &  1  &  0.2384  &  6.23  \\
\hline
RP\_20    &  1  &  0.2306  &  6.37  \\
\hline
RP\_24    &  1  &  0.2266  &  6.45  \\
\hline
RP\_28    &  1  &  0.2234  &  6.51  \\
\hline
$D_X(R)$   & 1   & 0.2166   & 6.64     \\
\hline
TCQ\_9        & 1   & 0.3945  & 4.04       \\
\hline
TCQ(opt)\_9   & 1   & 0.2793  & 5.54       \\
\hline
LC\_10        & 1   & 0.2529  & 5.97       \\
\hline
LC(opt)\_10    & 1   & 0.2495  & 6.03       \\
\hline
Pearlman\_10   & 1   & 0.3058  & 5.1456       \\
\hline
\hline
           & Rate(bits) & MSE    & SNR(dB) \\
\hline
\hline
RP\_24    &  2  &  0.0581  &  12.36  \\
\hline
$D_X(R)$   & 2   & 0.0538   & 12.69     \\
\hline
TCQ\_10        & 2   & 0.1194    & 9.23       \\
\hline
TCQ(opt)\_10   & 2   & 0.0755    & 11.22       \\
\hline
LC(opt)\_10    & 2   & 0.0668    & 11.75       \\
\hline
\hline
RP\_24    &  3  &  0.0152  &  18.18  \\
\hline
$D_X(R)$   & 3   & 0.0134   & 18.73     \\
\hline
TCQ\_11        & 3   & 0.0333    & 14.77      \\
\hline
TCQ(opt)\_11   & 3   & 0.0201    & 16.96      \\
\hline
\hline
RP\_24    &  4  &  0.0046  &  23.39  \\
\hline
$D_X(R)$   & 4   & 0.0033   & 24.79     \\
\hline
TCQ\_12        & 4   & 0.0089   & 20.53       \\
\hline
\end{tabular}
\end{center}
\caption{Laplacian Example}\label{table:laplace}
\vspace{-0.2in}
\end{table}

\begin{table}[htbp]
\begin{center}
\begin{tabular}{|c|c|c|c|}\hline
$y$ & 0.2  & 0.5  & 0.8 \\ \hline
$p_Y(y)$ & 0.368 & 0.264 & 0.368\\
\hline
\end{tabular}
\end{center}
\caption{Shannon Optimal Reproduction Distribution for the Uniform $(0,1)$ Source}\label{table:unifdist}
\end{table}

%%new table 11/19

\begin{figure}[htbp]
\begin{center}
 \includegraphics[width=3.6in]{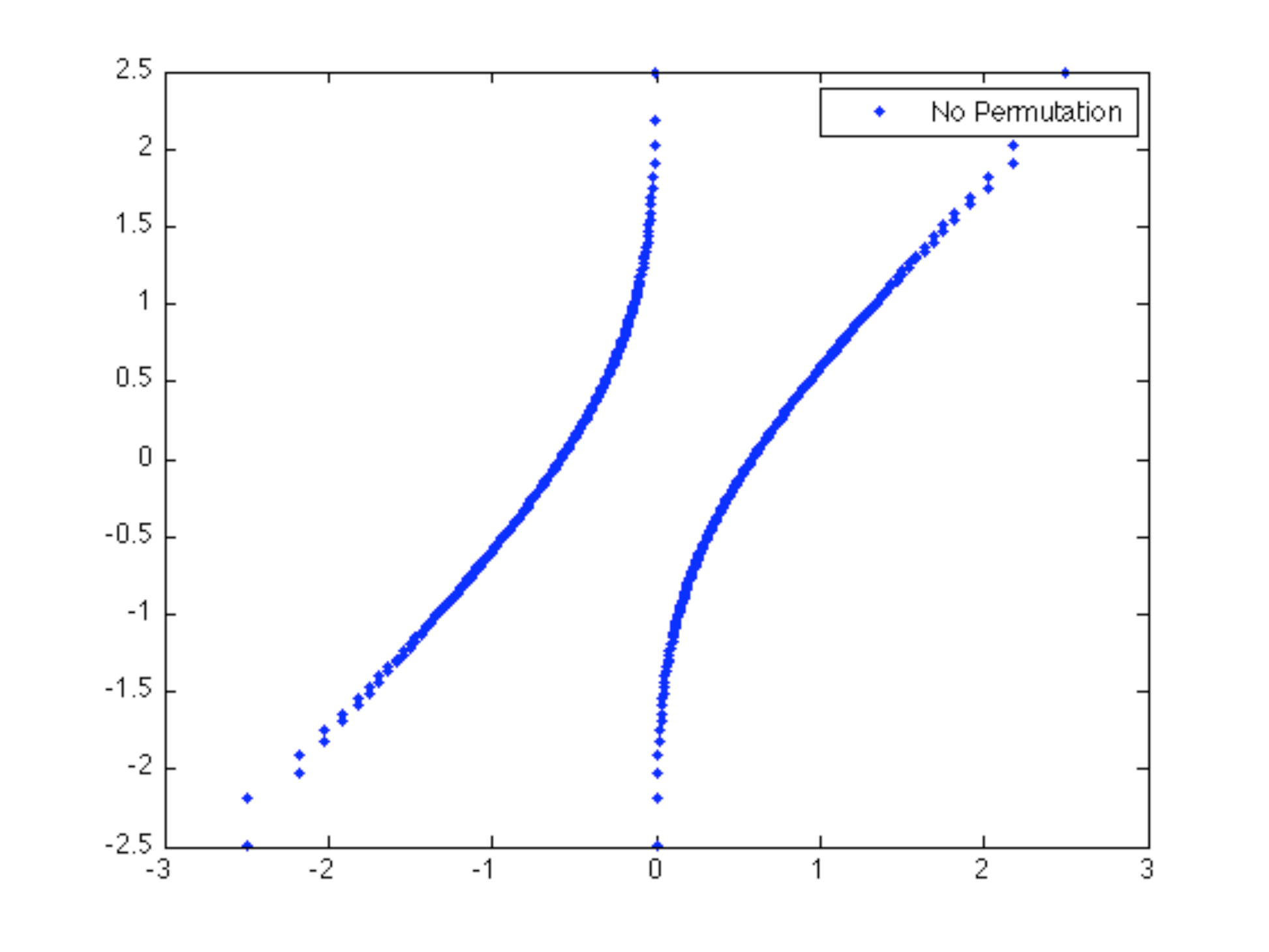}
  \includegraphics[width=3.6in]{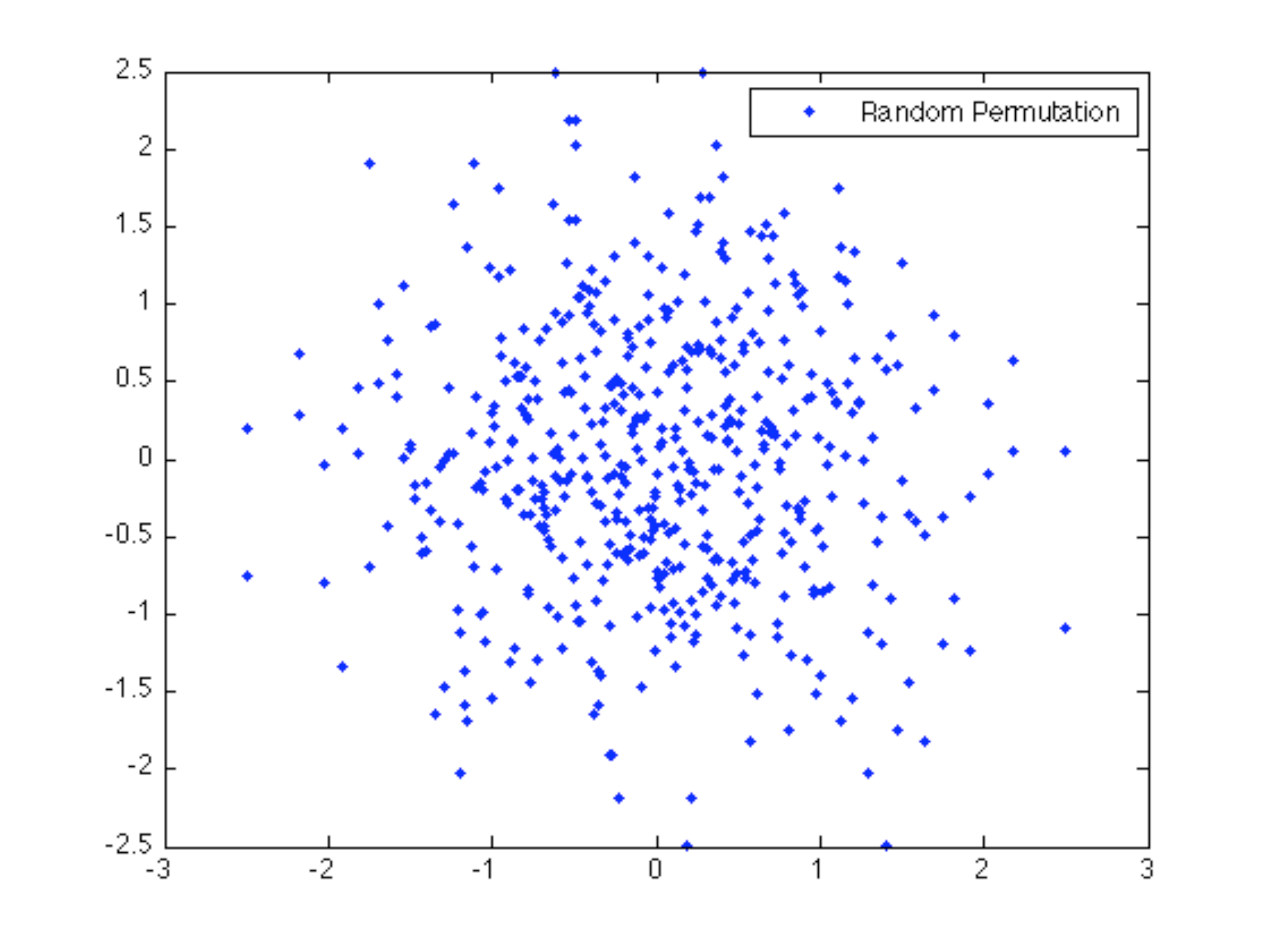}
\end{center}
\caption{Scatter plots of fake Gaussian 2-dimensional density: no permutation
and random permutation}\label{fig:higher}
\end{figure}

The effectiveness of the random permutation at forcing higher order distributions
to look more Gaussian is shown in Fig.~\ref{fig:higher}. The two dimensional scatter plot
for adjacent samples
with no permutation does not look Gaussian and is clearly highly correlated. When a randomly
chosen permutation is used, the plot looks like a 2D Gaussian sample. In both figures, the $x$ and $y$ axis are the value of the samples.

\begin{figure}[htbp]
\begin{center}
 \includegraphics[width=3.6in]{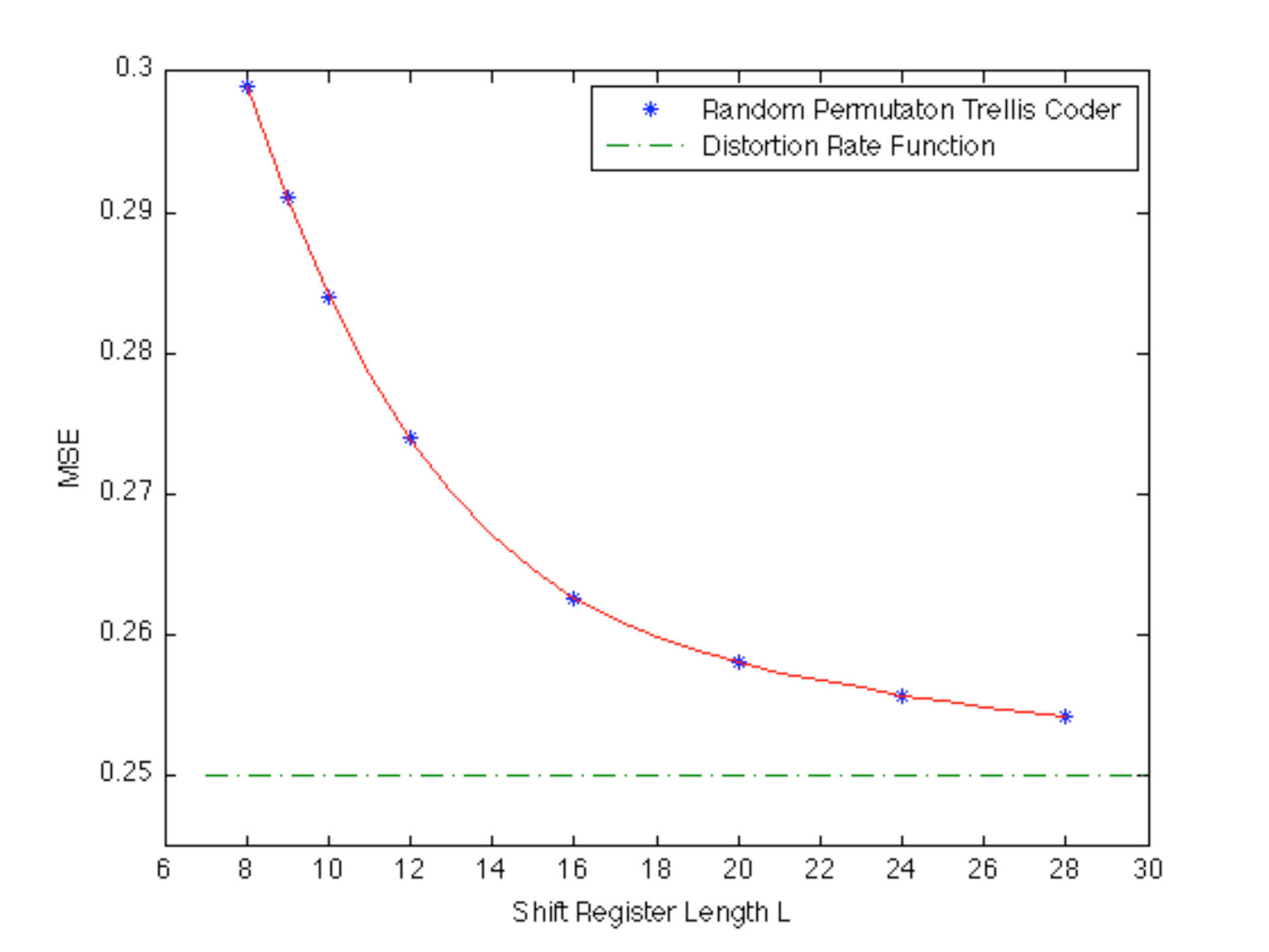}
\end{center}
\caption{Performance: 1 bit Gaussian}\label{fig:performance}
\end{figure}

Fig.~\ref{fig:performance}. shows the MSE of the random permutation trellis coder for IID Gaussian at $R=1$ with various shift register length. The performance has not yet shown to hit a plateau as shift register length increases.

The uniform IID source is of interest because it is simple, there is no exact formula
for the rate-distortion function with respect to mean-squared error and hence it must be found
by numerical means, and because one of the best compression algorithms, trellis-coded quantization (TCQ) 
is theoretically ideally matched to this example. So the example is an excellent one for demonstrating
some of the issues raised here and for comparison with other techniques.

The Rose algorithm yielded  a Shannon optimal distribution with
an alphabet of size 3 for $R=1$. %(fewer than Pearlman's choice of 4!). 
The points and their probabilities
are shown in Table~\ref{table:unifdist}.

\iffalse
\begin{table}[htbp]
\begin{center}
\begin{tabular}{|c|c|}\hline
$y$ & $p_Y(y)$ \\ \hline \hline
0.2       & 0.368\\
0.5        & 0.264\\
0.8        & 0.368\\
\hline
\end{tabular}
\end{center}
\caption{Shannon Optimal Reproduction Distribution for the Uniform $(0,1)$ Source}\label{table:unifdist}
\end{table}
\fi

Plugging the distribution into the random permutation trellis encoder led to a mapping $g$ of
(0 0.368) to 0.2, [0.368 0.632] to 0.5, and (0.632 1) to 0.8.

For the Laplacian source of variance $1$, 
the Rose algorithm yielded a Shannon optimal distribution with
an alphabet of size 9 for the 1 bit case. The 9 reproduction points
and their probabilities are listed in Table~\ref{table:laplacedist}.

%% Graphical illustration of a Laplacian source, the one bit
%% reproduction/simulation, and one bit scalar quantizer output is shown
%% in Figure~\ref{fig:fake_laplacian}.

\begin{table}[htbp]
\begin{center}
\begin{tabular}{|c|c|c|c|c|c|}\hline
$y$ & $\pm$ 4.6273 & $\pm$ 3.2828 & $\pm$ 2.1654 & $\pm$ 1.1063 &       0\\ \hline
$p_Y(y)$ & 0.0014  & 0.0065 & 0.0285 & 0.1266 & 0.6740 \\
\hline
\end{tabular}
\end{center}
\caption{Shannon Optimal Reproduction Distribution for the Laplacian Source}\label{table:laplacedist}
\end{table}

\iffalse
\begin{table}[htbp]
\begin{center}
\begin{tabular}{|c|c|}\hline
$y$ & $p_Y(y)$ \\ \hline \hline
-4.6273 & 0.0014 \\
-3.2828 & 0.0065 \\
-2.1654 & 0.0285 \\
-1.1063 & 0.1266 \\
      0 & 0.6740 \\
 1.1063 & 0.1266 \\
 2.1654 & 0.0285 \\
 3.2828 & 0.0065 \\
 4.6273 & 0.0014 \\
\hline
\end{tabular}
\end{center}

\caption{Shannon Optimal Reproduction Distribution for the Laplacian Source}\label{table:laplacedist}
\end{table}
\fi

For all three test sources --- Gaussian, uniform, and Laplacian ---  the performance
of the random permutation trellis source encoder is approaching the Shannon limit. 
Therefore, it is of interest to
estimate the entropy rate of the encoder output bit sequence, which should be
close to an IID equiprobable Bernoulli process
since an entropy rate near 1 is a necessary condition for 
approximate optimality~\cite{Gray:08,GrayL:09}.
A "plug-in"(or maximum-likelihood) estimator was used for this purpose.
The estimator uses the empirical probability of all words of a fixed length
in the sequence to estimate the entropy rate.
Bit sequences of length $10^6$ produced by encoding the Gaussian, uniform, and Laplacian sources with trellis encoder of shift register length $L=12$ were fed into the estimator,
the resulting entropy rate estimation ranges from $0.9993$ to $0.9995$.
For comparison, the estimator yielded entropy rate of 0.9998 for a randomly
generated bit sequence of the same length. %\clearpage

Eriksson et al.'s LC results for 1 bit at 512 states (equivalent to
shift register length 10) for Gaussian source is better than the
random permutation results for the same shift register length.
This is likely the result of their   exhaustive search
over all possible ways of labeling the branches within the constraint
of their  axioms. 
A similar approach to the random permutation code would be to search for the
permutation that produced the best results.
 Our results are from randomly chosen permutations, 
 so they reflect performance of the ensemble average (which we believe
 may eventually lead to a source 
 coding theorem using random coding ideas).
All permutations have the same  marginals, but some permutations 
will have better higher order distributions. 
%For example, no permutation at all, which is also a permutation, has very bad
%second order distributions as previously discussed.
Such an optimization is feasible only for small $L$.
We tested an optimization by exhaustion for $L=3$ and found that
the best MSE (SNR) was 0.3262 (4.8647), while the average MSE  (SNR) for all 
permutations was 0.3852 (4.1431). This demonstrates that the best permutation
can provide notable improvement over the average, but we have no efficient search
algorithm for finding optimum permutations.

\appendices
\section{Proof of Lemma \ref{lemma:GL}}
The encoded and decoded processes are both
stationary and ergodic since the original source is. 
From (\ref{eq:deltadbar}) and the source coding theorem,
\begin{eqnarray*}
D(f_n,g_n)&=&E[d(X_0,\hat{X}_0^{(n)})]\ge
 \dbar (\mu_X,\mu_{\hat{X}^{(n)}})\\
 &\ge&
 \infimum_{\nu: H(\nu )\le R}\dbar (\mu_X,\nu)=D_X(R).
 \end{eqnarray*}
The second inequality follows since stationary coding reduces entropy
rate,
and so $R   \ge {H}(U^{(n)}) \ge {H}(\hat{X}^{(n)})$. 
Since the leftmost term converges to the rightmost, the first equality
of the lemma is proved.

Standard  inequalities 
of information theory yield
\[
R  \ge {H}(U^{(n)}) \ge {H}(\hat{X}^{(n)}) \ge
{I}(X,\hat{X}^{(n)})
\ge
R_X (D(f_n,g_n))
\]
where the second inequality follows since mutual information rate is
bounded above by entropy rate, and the third  inequality follows from
the process definition of the Shannon rate-distortion
function~\cite{Gray:90}. %\cite{Marton:75,GrayNO:75}.  
Taking the limit as
$n\rightarrow\infty$, the rightmost term converges to $R$ since the
code sequence is asymptotically optimal (so that $D(f_n,g_n)\to
D_X(R)>0$) and the Shannon
rate-distortion function is a continuous function of its argument
(except possibly at $D=0$). Thus
$
\lim_{n\rightarrow\infty}{H}(U^{(n)}) =\lim_{n\rightarrow\infty}{H}(\hat{X}^{(n)}) =R.
$
proving the second equality of the lemma.

The final part requires Marton's inequality \cite{Marton:97}  relating
Ornstein's $\dbar$ distance and relative 
entropy when one of the processes is IID. 
Suppose that $\mu_U$ and $\mu_Z$ are stationary process distributions for two
processes with a common discrete alphabet
and that 
$\mu_{U^N}$ and $\mu_{Z^N}$ denote the finite dimensional distributions.
For any integer $N$ the  relative entropy or informational
divergence 
is defined by  
$$
H(\mu_{U^N} \| \mu_{Z^N} )=
\sum_{u^N} \mu_{U^N} (u^N)\log {\mu_{U^N}(u^N)\over \mu_{Z^N}(u^N)}.
$$

In our notation Marton's inequality states that
if $U$ is a stationary ergodic process and $Z$ is an IID process, then
$$
N^{-1} \mathcal{T}_0 (\mu_{U^N},\mu_{Z^N})
\le
 \left[ {\ln 2\over 2N} H(\mu_{U^N} \| \mu_{Z^N}))\right]^{1/2}.
 $$
Since $Z$ is an IID equiprobable process with alphabet size $2^R$, 
$$
N^{-1} \mathcal{T}_0 (\mu_{U^N},\mu_{Z^N})
\le
 \left[ {\ln 2\over 2N} ( NR-H(U^N))\right]^{1/2}
 $$
 and taking the limit as $N\rightarrow \infty$ yields (in view of
 property (\ref{eq:sup-limit}) of the $\dbar$ distance)
 $$
 \dbar_0(\mu_U,\mu_Z)\le 
  \left[ {\ln 2\over 2} ( R-H(U))\right]^{1/2}.
  $$
  Applying this to $U^{(n)}$ and taking the limit using the previous part of the lemma completes
  the proof.
\qed

\begin{lemma}\label{lemma:white2}\rm 
Let $\mu^N$ denote the $N$-fold product of a probability distribution
$\mu$ on the real line such that $\int x^2 d\mu(x) <\infty$. Assume
$\{\nu_n\}$ is a sequence of probability distribution on $\R^N$ such that
$%\[
 \lim_{n\to \infty} \mathcal{T}_2(\mu^N,\nu_n)=0.
$ %\]
If $Y^{(n)}_1,Y^{(n)}_2,\ldots, Y^{(n)}_N$ are random variables  with
joint distribution $\nu_n$, then for all $i\neq j$, 
\[
  \lim_{n\to \infty} E\bigl[\bigl(Y^{(n)}_i -E(Y^{(n)}_i)\bigr)\bigl(Y^{(n)}_j
    -E(Y^{(n)}_j)\bigr)\bigl]=0.
\]
\end{lemma}

\medskip

\emph{Proof.}\ 
The convergence of $\nu_n$ to $\mu^N$ in $\mathcal{T}_2$ distance
implies that there exist IID random variables  $Y_1,\ldots,Y_N$ with
common distribution $\mu$ and a sequence or $N$ random variables
$Y^{(n)}_1,Y^{(n)}_2,\ldots, Y^{(n)}_N$ with joint distribution
$\nu_n$, all defined on the same probability space, such that 
\begin{equation}
\label{eq:conv-l2}
\lim_{n\to \infty} E[(Y^{(n)}_i-Y_i)^2]=0, \quad i=1,\ldots,N.
\end{equation}
First note that this implies for all $i$
\begin{equation}
\label{eq:conv-secmoment}
\lim_{n\to \infty} E[(Y^{(n)}_i)^2]= E[Y_i^2]. 
\end{equation}  
Also,   $\lim\limits_{n\to \infty} E|Y^{(n)}_i - Y_i|=0$ (Cauchy-Schwarz),
so that for all $i$, 
\begin{equation}
\label{eq:conv-mean}
 \lim_{n\to \infty} E(Y^{(n)}_i) =E( Y_i). 
\end{equation}
Now the statement is direct convergence of the fact that in any inner
product space, the inner product is jointly continuous. To be more
concrete, letting $\langle X,Y \rangle = E(XY)$ and $\|X\|=[
  E(X^2)]^{1/2}$ for random variables $X$ and $Y$ with finite
second moment defined on this probability space, we have the bound
\begin{eqnarray*}
\lefteqn{
\bigl|\langle Y^{(n)}_i, Y^{(n)}_j\rangle - \langle Y_i, Y_j\rangle
\bigr|}\\ &\le & \bigl| \langle Y^{(n)}_i, Y^{(n)}_j-Y_j\rangle\bigr| +
\bigl| \langle Y^{(n)}_i- Y_i, Y_j\rangle\bigr| \\ 
&\le & \|Y^{(n)}_i\|\, \|Y^{(n)}_j-Y_j\| +\| Y^{(n)}_i- Y_i\|\, \| Y_j\|.
\end{eqnarray*}
Since $\|Y^{(n)}_i\|$ converges to $\|Y_i\|$ by
(\ref{eq:conv-secmoment}) and $\| Y^{(n)}_i- Y_i\|$ converges to zero by
(\ref{eq:conv-l2}), we obtain that $\langle Y^{(n)}_i,
Y^{(n)}_j\rangle$ converges to $ \langle Y_i, Y_j\rangle$, i.e,
\[
    \lim_{n\to \infty} E(Y^{(n)}_i Y^{(n)}_i)=  E(Y_iY_j)= E(Y_i)E(Y_j)
\]
since $Y_i$ and $Y_j$ are independent if $i\neq j$. This and
(\ref{eq:conv-mean}) imply the lemma statement. \qed

\section*{Acknowledgments}
The authors would like to thank an anonymous reviewer and the Associate
Editor for many constructive comments.

%\ifCLASSOPTIONcaptionsoff
%  \newpage
%\fi



%\bibliographystyle{IEEEtran}

\end{document}